\documentclass[preprint,3p,twocolumn]{elsarticle}

\usepackage{lineno,hyperref}
\modulolinenumbers[5]
\usepackage{color}
\usepackage{amsmath} 
\usepackage{cuted} 
\usepackage{subcaption} 
\usepackage{graphicx} 
\usepackage{bm}
\usepackage{amsfonts}
\usepackage{algorithm}
\usepackage[noend]{algpseudocode}

\usepackage{comment}

\journal{Medical Image Analysis}

\DeclareMathOperator*{\argmax}{argmax}

\biboptions{authoryear}

\begin{document}

\begin{frontmatter}

\title{Robust joint registration of multiple stains  and MRI for multimodal 3D histology reconstruction: Application to the Allen human brain atlas}

\author[affiliation_ucl]{Adri\`a Casamitjana}
\ead{a.casamitjana@ucl.ac.uk}

\author[affiliation_inria]{Marco Lorenzi}

\author[affiliation_ucl]{Sebastiano Ferraris}

\author[affiliation_ucl]{Loïc Peter}

\author[affiliation_kcl]{Marc Modat}

\author[affiliation_martinos]{\\Allison Stevens}

\author[affiliation_martinos,affiliation_ail,affiliation_hst]{Bruce Fischl}

\author[affiliation_kcl]{Tom Vercauteren}

\author[affiliation_ucl,affiliation_martinos,affiliation_ail]{Juan Eugenio Iglesias}

\cortext[mycorrespondingauthor]{Corresponding author}

\address[affiliation_ucl]{Center for Medical Image Computing, University College London, UK}

\address[affiliation_inria]{Université Côte d\' Azur, Inria, Epione Team, 06902 Sophia Antipolis, France}

\address[affiliation_kcl]{School of Biomedical Engineering {\normalfont \&} Imaging Sciences, King’s College London, UK}

\address[affiliation_martinos]{Martinos  Center  for  Biomedical  Imaging,  Massachusetts  General  Hospital  and  Harvard  Medical School, USA}

\address[affiliation_ail]{Computer  Science  and  Artificial  Intelligence  Laboratory,  Massachusetts  Institute  of  Technology, USA}

\address[affiliation_hst]{Program in Health Sciences and Technology, Massachusetts Institute of Technology, USA}

\begin{abstract}

Joint registration of a stack of 2D histological sections to recover 3D structure (``3D histology reconstruction'') finds application in areas such as atlas building and validation of \emph{in vivo} imaging. Straightforward pairwise registration of neighbouring sections yields smooth reconstructions but has well-known problems such as ``banana effect'' (straightening of curved structures) and ``z-shift'' (drift). While these problems can be alleviated with an external, linearly aligned reference (e.g., Magnetic Resonance (MR) images), registration is often inaccurate due to contrast differences and the strong nonlinear distortion of the tissue, including artefacts such as folds and tears. In this paper, we present a probabilistic model of spatial deformation that yields reconstructions for multiple histological stains that that are jointly smooth, robust to outliers, and follow the reference shape. The model relies on a spanning tree of latent transforms connecting all the sections and slices of the reference volume, and assumes that the registration between any pair of images can be see as a noisy version of the composition of (possibly inverted) latent transforms connecting the two images. Bayesian inference is used to compute the most likely latent transforms given a set of pairwise registrations between image pairs within and across modalities. We consider two likelihood models: Gaussian ($\ell_2$ norm, which can be minimised in closed form) and Laplacian ($\ell_1$ norm,  minimised with linear programming). Results on synthetic deformations on multiple MR modalities, show that our method can accurately and robustly register multiple contrasts even in the presence of outliers. The framework is used for accurate 3D reconstruction of two stains (Nissl and parvalbumin) from the Allen human brain atlas, showing its benefits  on real data with severe distortions. Moreover, we also provide the registration of the reconstructed volume to MNI space, bridging the gaps between two of the most widely used atlases in histology and MRI. The 3D reconstructed volumes and atlas registration can be downloaded from \url{https://openneuro.org/datasets/ds003590}. The code is freely available at \url{https://github.com/acasamitjana/3dhirest}.
\end{abstract}


\end{frontmatter}

\section{Introduction}

\subsection{Motivation}

Histology is the area of science concerned with microscopic exploration of tissue sections sampled from either a \emph{post mortem} specimen or biopsy tissue. After a tissue processing pipeline \citep{bancroft2008theory}, thin sections can be inspected under the microscope and digitised with a scanner. The most common histology pipeline consists of fixation, processing and embedding with a hardening material (e.g., wax) for sectioning. Thin sections from the wax-embedded tissue are cut using a microtome and mounted in glass slides for staining. Large specimens (e.g., a whole human brain) are typically first cut into several blocks, which are then  processed independently. 

Histological examination is the gold standard for many diagnostic protocols. Different staining procedures enable the visualisation of different microscopic structures. For example, the ubiquitous Haemotoxylin and Eosin (H\&E) \citep{chan2014wonderful} stains cell nuclei purple and cytoplasm pink while immunohistochemistry techniques selectively identify  antigens in cells \citep{ramos2005technical}). Stains are most often used in combination, e.g., to target different cell types in cancer diagnosis \citep{cooper2009feature} or to detect neuropathologies in neurodegenerative diseases  \citep{montine2012national}.


On addition to clinical pathology, histology has many applications in medical imaging, often in combination with $mm$-scale modalities like Magnetic Resonance Imaging (MRI): the former yields excellent contrast at the microscopic scale, while the latter provides larger-scale context, 3D structure and minimal distortion. One such application is the validation of \emph{in vivo} imaging techniques, like microstructure imaging  \citep{bourne2017apparatus} or mass spectrometry imaging \citep{thiele20142d}. In these applications, histology provides a gold standard for the underlying anatomy. 

Another successful application of combining $\mu m$-scale histology with $mm$-scale is the 3D modelling of anatomy at the microscopic level. In anatomy,  histology is often used as the basis for fine anatomical delineation as it provides detailed information about the size, shape and cell density of structures.
The $mm$-scale images are often used to inform the spatial registration of histological sections into a consistent 3D volume -- a problem known as 3D histology reconstruction \citep{pichat2018survey}. Sample applications include mammary glands \citep{shojaii2014reconstruction} or lungs \citep{rusu2015framework} in mice; brain tissue \citep{malandain2004fusion,stille20133d} or prostate \citep{gibson20133d} in humans.

In the specific case of human neuroimaging, histology has been combined with MRI to build  atlases. The most notable examples are the BigBrain project \citep{amunts2013bigbrain} and the Allen atlas \citep{ding2016comprehensive}, which used special whole-brain microtomes to section through a whole human brain specimen each (a single hemisphere in the Allen atlas), without needing to block the tissue. At the single structure level, \cite{yelnik2007three} used immunohistochemistry and MRI to build an atlas of the basal ganglia; \cite{adler2018characterizing} used a similar approach to build an atlas of the hippocampus; in previous work, we used  Nissl staining with thionin to build a probabilistic atlas of the human thalamus \citep{iglesias2018probabilistic}. \cite{krauth2010mean} also built a thalamic atlas, but used histological sections in different orientations (rather than a reference MRI) to solve the 3D reconstruction problem \citep{song20133d}. Many of these works integrate multiple stains to capture a richer characterisation of the microstructural organisation of tissue. For example, the Allen atlas comprises three staining procedures (Nissl stain and two antibodies for immunohistochemistry) to gather structural evidence from both cyto- and chemoarchitectural features \citep{ding2016comprehensive}. 

A central component of the works above is the 3D histology reconstruction. The invasive acquisition and the processing pipeline used in histology heavily distort the original shape of the tissue such that the 3D contextual information is lost and the spatial relationship between and within structures is broken. Therefore, image registration algorithms are required to recover the original 3D shape.
The tissue processing pipeline also produces a number of other artifacts, such as staining inconsistency, tears, folds, tissue loss, or air bubbles. These artifacts are often very difficult to model with deformation fields estimated with a registration algorithm, if not impossible (e.g., folding). Therefore, they are typically corrected with dedicated preprocessing methods \citep{pichat2018survey}.

In this work, we seek to achieve 3D reconstruction of multi-modality serial histology with a method that satisfies three main properties. First, producing spatially smooth reconstructions. Second, producing precise reconstructions, i.e., recovering shapes that accurately follow the underlying anatomy. And third, being robust against histological artefacts (e.g., folds or tears). We assume the availability of an external reference (e.g., an MRI scan, \citealt{annese2012importance}) that provides contextual information and enables unbiased reconstruction.

\subsection{Related work}
3D histology reconstruction without any additional shape information is an underconstrained problem. In this case, reconstruction is typically achieved by pairwise registration of adjacent slices. An important design choice in this approach is the reference slice, which is often chosen to be the one at the centre of the stack \citep{ourselin2001reconstructing} -- even though  automatic selection methods \citep{bagci2010automatic} have also been proposed. While these methods yield 3D reconstructions that are smooth (and thus visually pleasant), the lack of external guidance tends to straighten curved shapes (incurring the so-called ``banana effect'') and also leads to accumulation of errors along the stack (``z-shift'', \citealt{malandain2004fusion}). 

These problems can be mitigated with a reference volume providing information on the true shape and thus constraining the original reconstruction problem. 3D histology reconstruction is then split into a linear 3D registration problem between the reference volume and the stack of histological images, and a set of nonlinear 2D registration problems between section in the stack and the corresponding resampled slice from the reference volume. The problem can be addressed in an iterative fashion \citep{malandain2004fusion}. Other existing approaches attempt to solve the linear 3D and nonlinear 2D problems simultaneously, i.e., jointly optimising the registration similarity metric with respect to all linear and nonlinear parameters \citep{alic2011facilitating,yang2012mri}. While this approach is potentially more accurate, it also requires dedicated registration algorithms. 

Different methods have been used in the literature to initialise the stack, e.g., direct stacking of sections with alignment of their centres of mass \citep{goubran2013image} or, most commonly, pairwise registration of consecutive slices starting from the bottom \citep{ceritoglu2010large} or middle \citep{stille20133d} of the stack. To decrease the z-shift effect and avoid large error due to badly distorted slices, some approaches consider larger neighbourhoods when registering the histological stack. For example, \cite{yushkevich20063d} consider a 5-neighbourhood centred on the reference slice and use a graph theoretical approach to find the shortest path from every slice to a selected reference slice, increasing the robustness against poorly registered slices. The use of an undistorted intermediate modality may also facilitate the 3D alignment. For example, blockface photographs are sometimes taken previous to sectioning and thus do not present many of the artefacts caused by the histology processing pipeline \citep{amunts2013bigbrain}. 

If the linear 3D alignment is considered fixed (i.e., histological sections are linearly aligned to resampled slices of the reference volume), histology reconstruction reduces to a set of 2D in-plane registration problems. Naively, any intermodality registration model  could be used to align histology and the resampled slices of the reference volume (henceforth ``reference slices'') one at the time, producing an unbiased 3D reconstruction. However, intermodal nonlinear registration (typically using mutual information) is often difficult and inaccurate due to the artefacts discussed above, such as folding, tears, etc. \citep{jacobs1999registration}. When treating each histological section independently, these inaccuracies yield jagged reconstructions in the orthogonal planes.

To improve reconstruction continuity, sequential approaches in the literature consider not only the reference slice but also the adjacent histological sections in each 2D registration step \citep{adler2014histology,rusu2015framework,wirtz2004superfast}. Hence, each histological slice is deformed to simultaneously match their reference counterpart and neighbouring slices in an iterative fashion. However, these approaches are prone to getting stuck in local minima, are biased towards the choice of the initial slice, and may propagate correlated errors at each step.

Instead, joint refinement of  deformation fields provides a more robust alternative. For example, \cite{feuerstein2011reconstruction} define a 3D Markov random field between the control points of a set of precomputed 2D B-spline transforms and use discrete optimisation tools to globally minimise an energy function that encourages smoothness. Other \emph{ad hoc} approaches involve low-pass filtering of the deformation fields along the direction of the stack, e.g., with a Gaussian filter that smooths rigid transforms \citep{yushkevich20063d}. Similarly, \cite{malandain2004fusion} apply a Gaussian filter directly on linear transform parameters. Finally, \cite{casero2017transformation} define a new framework for linear histology reconstruction and provide theoretical equivalence with Gaussian filter smoothing. In our prior work \citep{iglesias2018model}, we defined a generative model over diffeomorphic deformation fields and used Bayesian inference to find a smooth solution along the stack direction. This framework yields 3D reconstructions that are simultaneously smooth and unbiased, i.e., without banana effect or z-shift.

\subsection{Contribution}
Here we present an extension of our previous conference article \citep{iglesias2018model} on 3D histology reconstruction with a reference volume. 
Our work seeks to model the nonlinear deformations that the tissue undergoes through the histology processing pipeline, i.e., we do not consider other artefacts that affect the topology of the tissue (e.g., folding or tearing) -- which, as mentioned above, can be addressed with dedicated algorithms \citep{pichat2018survey}.
The contributions with respect to the conference article are:  
\begin{itemize}
\item \textit{Multiple stains:} we extend the framework to joint reconstruction of multiple histological stains, explicitly encouraging the spatial alignment of all modalities. To the best of our knowledge, it is the first attempt at 3D histology reconstruction using multiple stains.
\item \textit{Robustness:} we greatly increase the robustness of our method against remaining artefacts by modelling the registration error with Laplacian distributions that penalise its $\ell_1$ norm, and solving the optimisation problem with linear programming. Moreover we use histology and reference masks on the generative model to further mitigate the impact of artefacts on the optimisation.
\item \textit{Computational efficiency:} we greatly increase the efficiency of our method by making approximations in the optimisation that have nearly no effect on the registration accuracy as well as by using modern deep learning registration techniques.
\end{itemize}

The rest of this paper is organised as follows. We introduce the proposed framework in Section~\ref{sec:methods}. Experiments on  two case studies  are described in Section~\ref{sec:experiments}. A dataset with synthetic deformations is used to thoroughly test and compare different variants of the methodology in Section~\ref{sec:synthetic}. A real-case scenario of 3D brain histology reconstruction is described in Section~\ref{sec:real}. Finally, Section~\ref{sec:discussion} discusses the results and concludes the article.

\section{Methods}
\label{sec:methods}

\subsection{Preliminaries}
\label{sec:preliminaries}
Consider a 3D histology reconstruction framework where different staining procedures are carried out and a reference volume is available (henceforth, we will assume this volume is an ex-vivo MRI scan).  Let $\{I^c_n(\bm{x})\}_{n=1,...,N}$ be a stack of $N$ histological sections of contrasts $c = 1,...,C$ (e.g., H\&E staining), defined on pixel locations $\bm{x}$ over a discrete 2D image domain $\Omega$. Paired sections of each contrast are cut a few microns apart and are thus assumed to be from the same tissue -- but with independent deformations. 

We further assume that the reference MRI volume has been linearly aligned to the stack, and resampled into the planes of the histological sections:  $\{I^0_n(\bm{x})\}_{n=1,...,N}$. Typically, a 3D rigid body transform is used between the MRI volume and one of the available contrasts (e.g., $c=1$). Then, 3D histology reconstruction amounts to solving a set of interdependent 2D registration problems. An initial 2D linear alignment between every section of the remaining contrasts ($c=2, \cdots, C$) and the correspondent MRI slices is subsequently computed. This registration not only solves the linear component of the problem, but also greatly simplifies the nonlinear part by bringing all the images to the same 2D coordinate frame (with equal pixel dimensions). We finally assume that all images (MRI and histology) have associated binary masks $M^c_n(\bm{x})$ discriminating tissue vs. background and obtained with  manual delineation or (semi-)automatic methods (e.g., \citealt{wang2016deep}). 

\begin{figure}[h]
    \centering
    \includegraphics[width=1.05\linewidth]{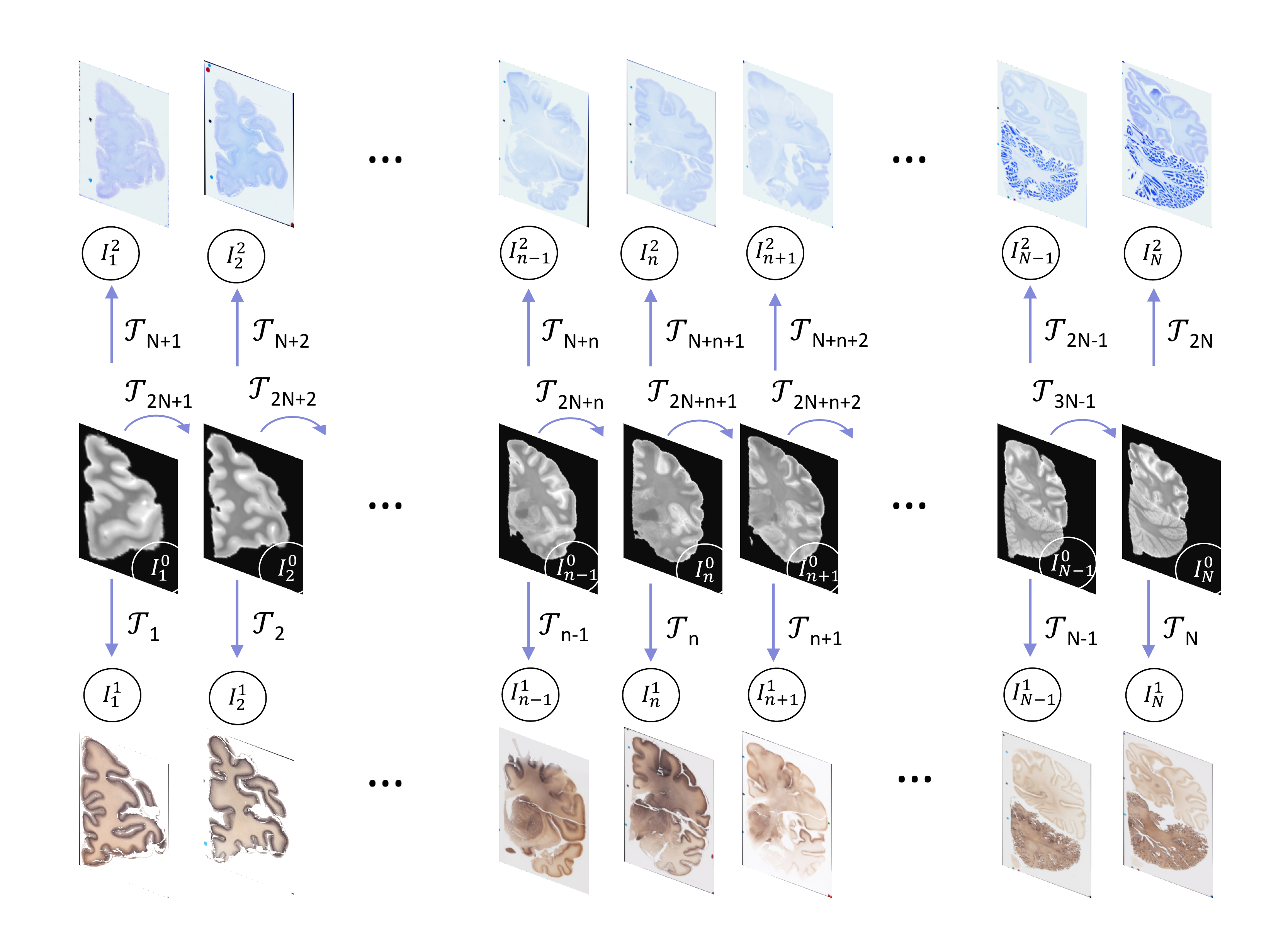}
    \caption{Our choice of  spanning tree for C=2. Each MRI slice is connected to the corresponding histological sections, as well as the immediate neighbour in the MRI stack.
    }
    \label{fig:spanning_tree}
\end{figure}

Furthermore, we consider all $(C+1)N$ images vertices in a graph $\mathcal{G}$, connected through a spanning tree with $L = N(C+1)-1 $ edges, such that any two images are connected by a unique path across the tree. The choice of spanning tree is irrelevant, as the algorithm presented below will guarantee convergence to the global optimum. Here we assume that the tree consists of $N-1$ edges connecting the MRI slices to their neighbours (i.e., $I_n^0$ to $I_{n+1}^0$, $\forall n<N$), as well a $C \times N$ edges connecting each MRI slice $I_n^0$ to the $C$ corresponding histological sections $I_n^1,\ldots,I_n^C$ (Fig.~\ref{fig:spanning_tree}).  

Associated with the edges of the spanning tree, we define a set of $L$ latent, noise-free, nonlinear, diffeomorphic transforms $\{\mathcal{T}_l(\bm{x})\}_{l=1,..., L}$. These transforms introduce directionality in the graph $\mathcal{G}$. As for the choice of spanning tree, the chosen criterion for defining the direction of the transforms does not affect the results of the algorithm. Here, we assume that the transforms point from $I_n^0$ to $I_n^c$, $\forall n,c$, and from $I_n^0$ to $I_{n+1}^0$, $\forall n<N$ (Fig.~\ref{fig:spanning_tree}). Since these transforms are assumed to be diffeomorphic and thus invertible, one can obtain the latent transform connecting any two images in $\mathcal{G}$ by composing a subset of (possibly inverted) transforms in $\{\mathcal{T}_l(\bm{x})\}$.

\begin{figure}[h!]
    \centering
    \includegraphics[width=0.45\textwidth]{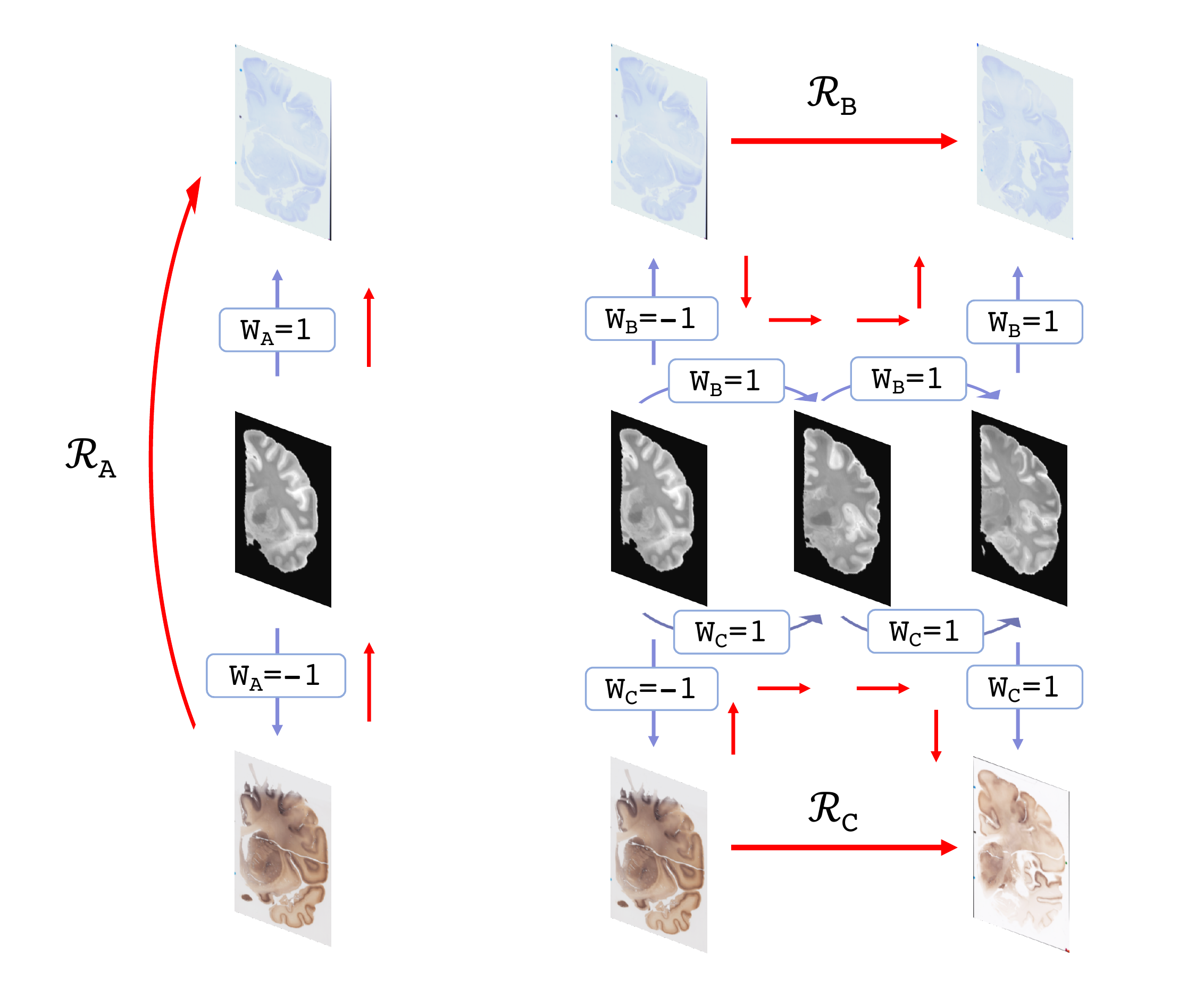}
    \caption{Example of three observations. $\mathcal{R}_A$: an intermodality registration between histological contrasts. $\mathcal{R}_B, \mathcal{R}_C$: intramodality registration within contrasts in the tree leaves. Blue and red arrows indicate the direction of the velocity fields from the spanning tree definition and observation path, respectively. Weight values are +1/-1 when blue and red arrows follow the same/opposite direction.}
    \label{fig:registration_example}
\end{figure}

Finally, we consider a set of $K \geq L$ nonlinear diffeomorphic transforms between pairs of images in $\mathcal{G}$, estimated with a diffeomorphic registration algorithm (e.g., \citealt{avants2008symmetric,modat2012parametric,dalca2018unsupervised,ashburner2007fast,vercauteren2008symmetric}): $\{\mathcal{R}_k(\bm{x})\}_{k=1,..., K}$. Every $\mathcal{R}_k$ can be seen as a noisy version of a composition of transforms in $\{\mathcal{T}_l\}$ and their inverses. We use a $K \times L$ matrix $\bm{W}$ to encode the transforms in $\{\mathcal{T}_l\}$ that each $\mathcal{R}_k$ traverses, as follows:  
\begin{itemize}
    \item $W_{kl}=1$ if $\mathcal{T}_l$ is on the path of $\mathcal{R}_k$,
    \item $W_{kl}=-1$ if $\mathcal{T}^{-1}_l$ is on the path of $\mathcal{R}_k$, and
    \item $W_{kl}=0$ otherwise.
\end{itemize}
Even though $\mathcal{R}_k$ can in principle connect any pair of nodes, one would normally only register images that are not too far in the graph. In practice, we compute registrations between images at the same level in the stack (MRI to histology, as well as between different histological stains), as well as between every image (MRI or histology) and its nearest neighbours in the stack (Fig.~\ref{fig:registration_example}). It is the goal of our method to infer the most likely underlying  $\{\mathcal{T}_l\}$ given the observed $\{\mathcal{R}_k\}$, as explained in the following section.  

\subsection{Probabilistic modelling and Bayesian inference}
\label{sec:model}

In this work, we assume that the latent transforms $\{\mathcal{T}_l\}$ and the observed registrations $\{\mathcal{R}_k\}$ are explained by a probabilistic generative model, such that 3D histology reconstruction can be posed as a Bayesian inference problem:  given the $K \geq L$ observed noisy registrations, what is the most likely set of $L$ underlying transforms (i.e., those in the spanning tree) that gave rise to them?

The proposed probabilistic model relies on a number of key assumptions:
\begin{itemize}
    \item The observed registrations $\{\mathcal{R}_k\}$ are conditionally independent, given the latent transforms $\{\mathcal{T}_l\}$.
    \item The likelihood of each registration $\mathcal{R}_k$ is parameterised by a set of parameters $\bm{\theta}$, for which we do not make any prior assumptions, i.e., $p(\bm{\theta}) \propto 1$.
    \item We do not make any prior assumptions on the distribution of the latent transforms $\{\mathcal{T}_l\}$, i.e., p($\{\mathcal{T}_l\})\propto 1$.
\end{itemize}

Under these assumptions, the generative model describing the joint probability distribution of the latent transforms, the likelihood parameters and the observed registrations is:
\begin{align}
    p(\{\mathcal{T}_l\}, & \{\mathcal{R}_k\}, \bm{\theta}) \nonumber \\
    & = p(\{\mathcal{T}_l\}) p(\bm{\theta}) \prod_{k=1}^K p(\mathcal{R}_k | \{\mathcal{T}_l\}, \bm{\theta}; W_{k,:})  \nonumber \\
    & \propto \prod_{k=1}^K p(\mathcal{R}_k | \{\mathcal{T}_l\}, \bm{\theta}; W_{k,:}),  \label{eq:generative}
\end{align}
where we have dropped the dependency on $\bm{x}$ for simplicity.

Within this framework, one can compute the most likely 3D reconstruction by finding the most likely latent transforms that bring the histological sections into alignment; we note that this is a subset of $\{\mathcal{T}_l\}$, including only the transforms between histology and MRI (subset $S_1$); the transforms between reference MRI slices (subset $S_2$) are not needed for the 3D histology reconstruction. In a fully Bayesian formulation, solving this problem requires marginalising over all the variables we are not seeking to optimise, including the likelihood parameters and the subset of latent transforms connecting the reference MRI slices:
\begin{small}
\begin{align}
    &\{\hat{\mathcal{T}}_l\}_{l\in S_1}  = \argmax_{\{\mathcal{T}_l\}_{l\in S_1} } p(\{\mathcal{T}_l\}_{l\in S_1} | \{\mathcal{R}_k\}) \nonumber \\
    & = \argmax_{\{\mathcal{T}_l\}_{l\in S_1} } \int p(\{\mathcal{T}_l\}, \bm{\theta} | \{\mathcal{R}_k\}) d\bm{\theta}  \prod_{l\in S_2}d\mathcal{T}_l  . \label{eq:intractable} 
\end{align}
\end{small}

Equation~\ref{eq:intractable} is often intractable due to the integral over transforms and likelihood parameters. Instead, we optimise the joint probability of all transforms (both subsets) and likelihood parameters:
\begin{small}
\begin{align}
    \{\hat{\mathcal{T}}_l\}, \hat{\bm{\theta}}  & = \argmax_{\{\mathcal{T}_l\}, \bm{\theta} } p(\{\mathcal{T}_l\}, \bm{\theta} | \{\mathcal{R}_k\}) \nonumber \\
    & = \argmax_{\{\mathcal{T}_l\}, \bm{\theta} } p(\{\mathcal{T}_l\}, \bm{\theta}, \{\mathcal{R}_k\}) \nonumber \\
    & = \argmax_{\{\mathcal{T}_l\}, \bm{\theta} } \prod_{k=1}^K p(\mathcal{R}_k | \{\mathcal{T}_l\}, \bm{\theta}; W) \nonumber \\
    & = \argmax_{\{\mathcal{T}_l\}, \bm{\theta} } \sum_{k=1}^K \log p(\mathcal{R}_k | \{\mathcal{T}_l\}, \bm{\theta}; W).  \label{eq:approximate}
\end{align}
\end{small}

\subsection{Model instantiation}
\label{sec:model_instantiation}

There are two main design choices in our model: the representation for the spatial transforms $\{\mathcal{T}_l\},\{\mathcal{R}_k\}$, and the shape of the likelihood $p(\mathcal{R}_k | \{\mathcal{T}_l\}, \bm{\theta}; W)$.

\subsubsection{Model for spatial transforms}
\label{sec:model_transforms}

We choose the Log-Euclidean framework to parameterise diffeomorphisms in the Lie group of stationary velocity fields (SVFs, \citealt{arsigny2006log}). Let $\{\bm{R}_k(\bm{x})\}$ and $\{\bm{T}_l(\bm{x})\}$ be the SVF infinitesimal generators whose integration using Lie exponentials result in the corresponding diffeomorphisms $\mathcal{R}_k(\bm{x}) = \exp[\bm{R}_k(\bm{x})]$ and $\mathcal{T}_l(\bm{x}) = \exp[\bm{T}_l(\bm{x})]$. For fast computation of exponentials we use the scaling-and-squaring approach \citep{arsigny2006log}. From Lie group manifolds, two relevant properties are derived. First, the inverse of a transform is equivalent to its negation in the log-space:
$$
\mathcal{T}^{-1}_l(\bm{x}) = \exp[-\bm{T}_l(\bm{x})];
$$
and second, in a scenario of small deformations, the composition of  transforms can be approximated by truncating the Baker-Campbell-Hausdorff series at its first term \citep{vercauteren2008symmetric}:
$$\mathcal{T}_l(\bm{x}) \circ \mathcal{T}_{l'}(\bm{x}) \approx \exp[\bm{T}_l(\bm{x}) + \bm{T}_{l'}(\bm{x})].
$$
These two properties greatly simplify evaluation of the likelihood terms described below.



\subsubsection{Likelihood models}

We consider two different likelihood models, Gaussian and Laplacian. Both of them are based on two key assumptions. The first assumption is statistical independence across spatial locations and also between the horizontal and vertical components of the transforms. We note that, in spite of such spatial independence, the smoothness of the solution will  be guaranteed for two reasons: the observed registrations are often spatially smooth (leading to smooth solutions for $\{\mathcal{T}_l\}$), and the fact that we only solve the problem at a sparse set of control points, as explained in Section~\ref{sec:optimisation_details} below.
The second assumption is that each observed registration is a noisy version of the ``true'' underlying transform, which is a composition of a subset of the hidden transforms, possibly inverted, as specified by the matrix $\bm{W}$ (as explained in Section~\ref{sec:preliminaries} above). 

Let $R_k^{\xi_1}(\bm{x})$ and $R_k^{\xi_2}(\bm{x})$ be the horizontal ($\xi_1$) and vertical ($\xi_2$) components of the SVF of registration $k$ at $\bm{x}$, which we group into two $K\times 1$ vectors $\bm{R}^{\xi_j}(\bm{x}) =[R_1^{\xi_j}(\bm{x}),\ldots,R_K^{\xi_j}(\bm{x})]^T$, with $j\in\{1,2\}$.
In a similar fashion, let $T_l^{\xi_1}(\bm{x})$ and $T_l^{\xi_2}(\bm{x})$ be the two components of the SVF of the $l^{th}$ latent transform, grouped into two $L\times 1$ vectors $\bm{T}^{\xi_j}(\bm{x})=[T_1^{\xi_j}(\bm{x}),\ldots,T_L^{\xi_j}(\bm{x})]^T$, with $j\in\{1,2\}$. The model of spatial transforms in Section~\ref{sec:model_transforms} enables us to write:
$$
    \bm{R}^{\xi_j}(\bm{x}) = \bm{W} \bm{T}^{\xi_j}(\bm{x}) + \bm{\zeta}^{\xi_j}(\bm{x}), \quad \text{with}\ j\in\{1,2\},
$$
where $\bm{\zeta}^{\xi_1}(\bm{x})$ and  $\bm{\zeta}^{\xi_2}(\bm{x})$ are $K\times 1$ vectors with the horizontal and vertical components of  errors in the SVFs of the $K$ registrations at $\bm{x}$. 
The statistical distribution of the error $\bm{\zeta}$ (Gaussian or Laplacian) will shape the likelihood model, as described next.

\paragraph{Gaussian} 
The Gaussian model was presented in our previous conference article \citep{iglesias2018model}. In short, it assumes that each error $\zeta_k^{\xi_j}(\bm{x})$ is independent from the others and follows a Gaussian distribution with zero mean and variance  $\sigma^2_k$, such that the likelihood is:
\begin{equation}
    \bm{R}^{\xi_j}(\bm{x}) \sim \mathcal{N}\left(  \bm{W} \bm{T}^{\xi_j}(\bm{x}), \text{diag}[\sigma^2_k] \right). 
    \label{eq:gaussian_likelihood}
\end{equation}
We use a model for the variances $\{\sigma^2_k\}$, which explicitly assumes that errors are larger when registering across modalities, or when registering slices or sections further apart. Specifically, the variance of a registration is assumed to be a linear combination of inter- and intra-modal variances:
$$
\sigma^2_k = c_k \sigma^2_{inter} + \sum_{c=0}^C d_{k,c} \sigma^2_c,
$$
where $c_k \in \{0,1\}$ is an indicator variable which is equal to one if the registration is across modalities (e.g., MRI to histology, or between different histological stains), and zero otherwise; $d_{k,c} \geq 0$ is the separation between the registered slices or sections along the stack for the intramodal registration of the $c$-th constrast, and zero otherwise; and $\sigma^2_c,\sigma^2_c$ are model parameters that need to be estimated, i.e., $\bm{\theta}=[\sigma^2_{\text{inter}},\sigma^2_{c=0}, \cdots, \sigma^2_{c=C}]^T$.

\paragraph{Laplacian} 
In spite of the model for the variances, the Gaussian likelihood is sensitive to outliers in the registration. Such outliers occur frequently in histology due to common artifacts such as folding or tears. As an alternative, we propose a Laplacian model penalising the absolute value of the errors, i.e., the $\ell_1$ norm:
\begin{equation}
    R_k^{\xi_j}(\bm{x}) \sim \text{Laplace}\left(  \bm{W} \bm{T}^{\xi_j}(\bm{x}), b \right),
    \label{eq:laplacian_likelihood}
\end{equation}
where $b$ is the scaling parameter of the Laplace distribution, which we consider constant across registrations; this assumption enables us to solve a single linear program per location during inference -- rather than iteratively solving multiple linear programs. Moreover, modelling the dispersion of each registration separately as in the Gaussian case is not as important, due to the robustness of the $\ell_1$ norm against outliers. Therefore, $b$ is the only model parameter, i.e., $\bm{\theta}=[b]$.

\subsection{Inference algorithms}

Following the general inference framework in Section~\ref{sec:model} and the design choices  in Section~\ref{sec:model_instantiation}, we now present two specific algorithms to solve the inference problem for the two proposed likelihood models.

\subsubsection{Gaussian}

Substituting the Gaussian likelihood from Equation~\ref{eq:gaussian_likelihood} into the maximisation problem from Equation~\ref{eq:approximate} and switching signs, we obtain the following cost function \citep{iglesias2018model}:
\begin{align}
C_{\ell_2}[\bm{T}&^{\xi_1}  (\bm{x}),\bm{T}^{\xi_2}(\bm{x}),\sigma^2_c,\sigma^2_d)] = \nonumber \\
& |\Omega| \sum_{k=1}^K \log [2\pi (c_k \sigma^2_c + d_k \sigma_d^2)] \nonumber \\
 + & \sum_{j=1}^2 \sum_{k=1}^K \sum_{\bm{x}\in\Omega} \frac{[R_k^{\xi_j}(\bm{x})-\sum_{l=1}^L W_{kl} T_l^{\xi_j}(\bm{x})]^2}{2(c_k \sigma^2_c + d_k \sigma_d^2)}. \label{eq:problemL2} 
\end{align}

We use coordinate descent to solve this minimisation problem, alternately optimising for $\{\bm{T}^{\xi_j}\}_{j=1,2}$ and for $\bm{\theta}=[\sigma_c^2,\sigma_d^2]^T$, with the other fixed. For a constant $\bm{\theta}$, Equation~\ref{eq:problemL2} becomes a simple weighted least squares problem, with a closed-form solution given by:
\begin{equation}
T_l^{\xi_j}(\bm{x}) = \sum_{k=1}^K Z_{lk} R_k^{\xi_j}(\bm{x}),
\label{eq:latent_update}
\end{equation}
for the two spatial coordinates $\xi_j$, $j={1,2}$. The regression matrix $\bm{Z}$ is by:
$$
\bm{Z} = [\bm{W}^T \text{diag}(1/\sigma_k^2)  \bm{W}]^{-1} \bm{W}^T \text{diag}(1/\sigma_k^2).
$$

With the hidden transforms fixed, there is no closed-form expression for $\sigma_c^2$ and $\sigma_d^2$. However, Equation~\ref{eq:problemL2} becomes a smooth function of two variables that can be easily and quickly minimised with numerical methods, e.g., conjugate gradient \citep{shewchuk1994introduction} or BFGS \citep{liu1989limited}.

\subsubsection{Laplacian}
\label{sec:laplacian}

Substituting the Laplacian likelihood from Equation~\ref{eq:gaussian_likelihood} into  Equation~\ref{eq:approximate}, we obtain the following objective function:
\begin{align}
\mathcal{O}_{\ell_1} =  & - 2 K |\Omega| \log (2b) \nonumber \\
  & - \frac{1}{b} \sum_{j=1}^2\sum_{k=1}^K \sum_{\bm{x}\in\Omega}
 |R_k^{\xi_j}(\bm{x})-\sum_{l=1}^L W_{kl} T_l^{\xi_j}(\bm{x})|. \label{eq:objectiveL1} 
\end{align}
Since the values of the optimal latent transforms that minimise Equation~\ref{eq:objectiveL1} do not depend on the model parameter $b$, we can remove the terms related to $b$ and switch signs to obtain the following cost function:
\begin{align}
C_{\ell_1}[\bm{T}&^{\xi_1}  (\bm{x}),\bm{T}^{\xi_2}(\bm{x})] = \nonumber \\
&  \sum_{j=1}^2 \sum_{k=1}^K \sum_{\bm{x}\in\Omega}
 |R_k^{\xi_1}(\bm{x})-\sum_{l=1}^L W_{kl} T_l^{\xi_1}(\bm{x})|, \label{eq:problemL1} 
\end{align}
which can be solved one spatial location $\bm{x}$ and  direction (horizontal $\xi_1$ or vertical $\xi_2$) at the time. Crucially, the minimisation of Equation~\ref{eq:problemL1}
 can be rewritten as a linear program in standard form as follows:
\[\arraycolsep=1.4pt\def\arraystretch{1.5}
\begin{array}{rr@{}ll}
\label{eq:linear_program}
\text{minimize}  & \displaystyle \bm{c}^{T}\bm{y}    & \\
\text{s. t.}     & \displaystyle \bm{A}_1^{T} & \bm{y} \leq -\bm{R}^{\xi_j}(x),\\
                 & \displaystyle \bm{A}_2^{T} & \bm{y} \leq \bm{R}^{\xi_j}(x),\\
\end{array}
\]
where:
\begin{itemize}
\item $\bm{y} = [D^{\xi_j}_1(\bm{x}),...,D^{\xi_j}_K(\bm{x}),T^{\xi_j}_1(\bm{x}),..., T^{\xi_j}_{L}(\bm{x})]^T$ is a $(K+L)\times 1$ vector concatenating the K absolute deviations , $D^{\xi_j}_k(\bm{x})$ (defined below), and the latent transforms to estimate, $T^{\xi_j}_l(\bm{x})$. 
\item $\bm{c} = [\bm{1}_K^T, \bm{0}_L^T]^T$, where $\bm{1}_K$ and $\bm{0}_L$ are the all-one and all-zero vectors with dimensions $K\times 1$ and $L \times 1$, respectively. 
\item $\bm{A}_1 = [-\bm{I}_K, -\bm{W}]$, where $\bm{I}_K$ is the $K \times K$ identity matrix.
\item $\bm{A}_2 = [-\bm{I}_K, \bm{W}]$.
\end{itemize}

Since the vector $\bm{c}$ zeroes out the second part of $\bm{y}$, the objective function of this linear program and Equation~\ref{eq:problemL2} are identical. The inequality constraints effectively force the deviations $D^{\xi_j}_K(\bm{x})$ to be positive and equal to:
$$
D^{\xi_j}_K(\bm{x}) = |R_k^{\xi_j}(\bm{x})-\sum_{l=1}^L W_{kl} T_l^{\xi_j}(\bm{x})|.
$$
Therefore, the linear program is equivalent to the problem of minimising $\mathcal{C}_{\ell_1}$ in Equation~\ref{eq:problemL1}, with the difference that we can now use well-established linear programming algorithms to obtain the solution -- which is simply the second part (last $L$ elements) of the optimal $\bm{y}$. 

\subsection{Spatially-varying subgraphs}
So far we have assumed any location in the image domain ($\bm{x} \in \Omega$) is modelled with the same static graph structure, encoded in the matrix $\bm{W}$. However, it is desirable to model in the graph image features and artefacts that may influence registration. For example, one may want to remove edges from $\mathcal{G}$ when modelling pixels far away from the masks $\{M_n^c\}$, which cannot be mapped reliably across images, or edges modelling pixels with artefacts such as folds and tears. 

To tackle this problem, we propose to build spatially-varying subgraphs of $\mathcal{G}$ for each spatial location $\bm{x} \in \Omega$, which we represent as  $\mathcal{G}(\bm{x})$.  We also build companion matrices $\bm{W}=\bm{W}(\bm{x})$ that encode the relationship between the registrations and the hidden transforms at each location, considering the structures of $\mathcal{G}(\bm{x})$. These subgraphs $\mathcal{G}(\bm{x})$ contain a subset of the edges in $\mathcal{G}$ and 
are built by simply removing from  $\mathcal{G}$ all connections going from $I_n^c(\bm{x})$ to $I_{n'}^{c'}(\bm{x})$, for which no tissue is present in the mask or the source  image at location $\bm{x}$, i.e., when $M_n^c(\bm{x})=0$.

This approach easily accommodates the algorithm to cases for which we have irregular graphs; e.g., histological contrasts with different number of sections or missing correspondence between contrasts.

\subsection{Registration networks}
\label{sec:registration_networks}
A learning-based strategy inspired by \citep{balakrishnan2019voxelmorph} is used to compute the SVF maps $\{\bm{R}_k(\bm{x})\}$. It consists of a backbone network (U-Net type, \citealt{cciccek20163d}) that produces a low-resolution SVF at 1/8 of the original resolution.  Then, a rescaling layer with linear interpolation is used to get a full resolution velocity field. ``Scaling and squaring''  \citep{arsigny2006log} is used to integrate the SVFs and compute the deformation fields $\{\mathcal{R}_k(\bm{x})\}$ used for spatial alignment. The inverse deformation field is computed by integrating the negated SVFs and used to induce symmetry in the training by evaluating the cost function at both reference and target image spaces. 
A local normalised cross-correlation and a smoothness regularisation term constraining the spatial gradient of the deformation are used as the composite loss function in intramodality networks. For intermodality registration, we use our synthesis based method presented in \cite{casamitjana2021synthbyreg}. In short, this method uses a registration loss for weakly supervised image translation between domains where corresponding images exist but the spatial correspondence is not known (i.e., as between histology and MRI). The registration loss is complemented with a structure preserving constraint based on contrastive learning, and the algorithms produces an inter-modality registration as by-product.

The learning approach is used to train a total of $C\times(C+1)/2$ registration networks on the entire dataset, one for every possible pair of contrasts. Intermodality networks are trained with randomly selected pairs of matching images $I_{n}^{c}$, $I_{n}^{c'}$ at every minibatch (i.e., with fixed $n$, and $c\neq c'$). Intramodality networks use randomly selected pairs of images, constrained to be within a maximum number of slices -- in practice, we use 4, i.e., $I_{n}^{c}$, $I_{n'}^{c}$, with fixed $c$ and $|n-n'|\leq 4$.
In order to increase the generalisation ability of the models, all networks use an augmentation scheme whereby images are spatially deformed at every minibatch with a random, smooth, nonlinear displacement field, obtained by linear interpolation of a low resolution grid of control points ($9\times 9$) whose strength depends on the image resolution.

\subsection{Implementation details and summary of the algorithm}

\subsubsection{Structure of the observational graph}
\label{sec:graph_structure}
\begin{figure}[h]
    \centering
    \includegraphics[width=\linewidth]{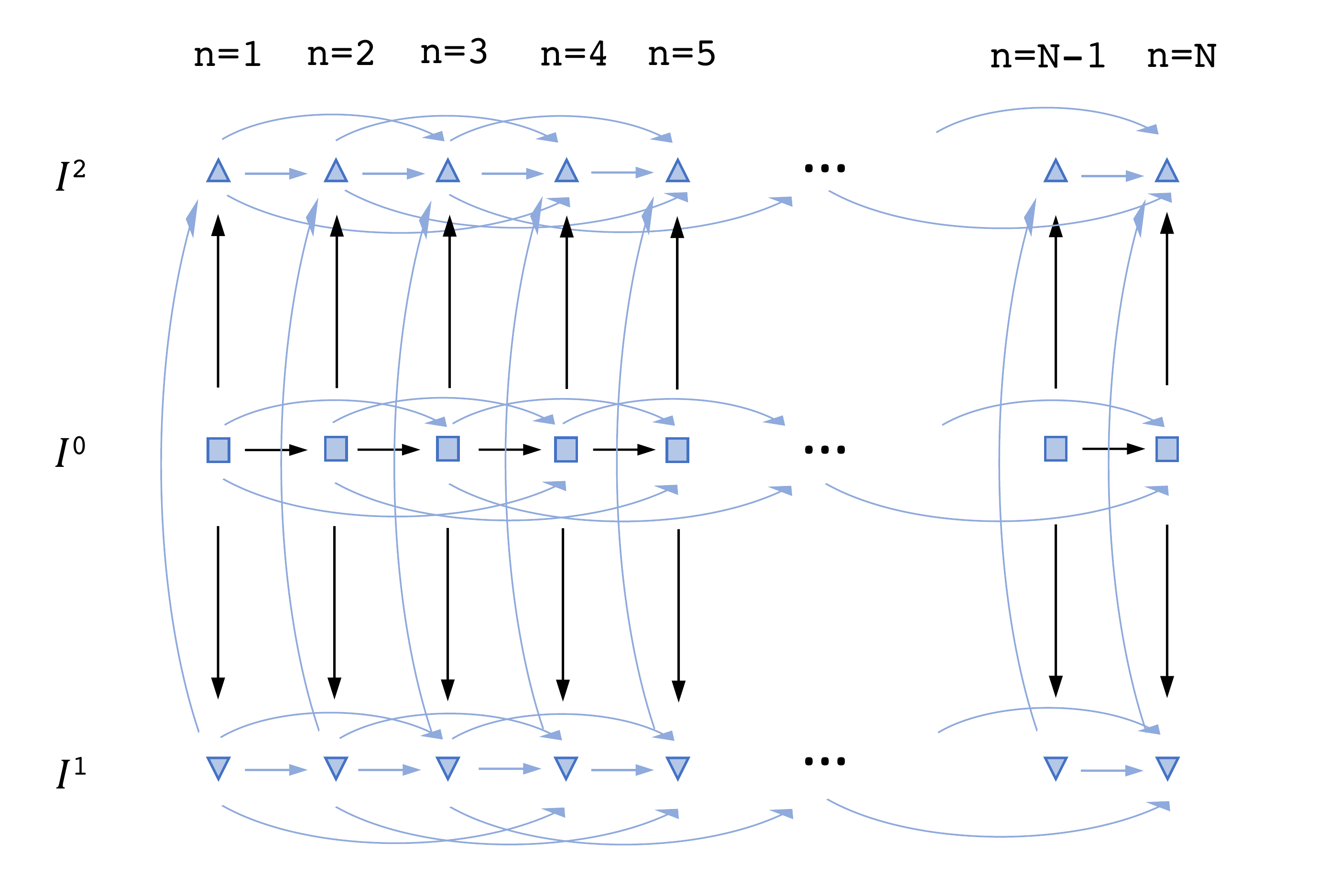}
    \caption{Example of the observational graph used with $C=2$ and P=3. 
    Arrows show all computed registrations in the graph with their corresponding direction and the spanning tree is highlighted in black.}
    \label{fig:observation_graph}
\end{figure}

As explained in Section~\ref{sec:preliminaries}, we compute $K$ registrations between pairs of images in the graph. First, we compute intermodality registrations between all pairs of corresponding slices from different contrasts, i.e., between $I_n^c$ and $I_n^{c'}$, with $c\neq c'$. And second, we compute intramodality registrations between every pair of images of the same contrast, that are not more than $P$ slices apart, i.e., between $I_n^c$ and $I_{n'}^c$, with $|n-n'|\leq P$. The maximum separation $P$ controls the smoothness of the solution (higher values of $P$ yield smoother reconstruction but also more banana effect). An example of observational graph is shown in Figure~\ref{fig:observation_graph}.

\subsubsection{Optimisation details}
\label{sec:optimisation_details}
Since registration yields smoothly varying SVF maps (upscaled from 1/8 resolution, as explained in Section~\ref{sec:registration_networks}), we greatly reduce the computational requirements of our method by running the proposed inference algorithm on low-resolution SVFs. As in Section~\ref{sec:registration_networks}, we use linear interpolation to upsample the refined SVFs to the full resolution and use scaling and squaring to compute the final deformation fields. 

In terms of  optimisation approaches, we use different methods for the two likelihood models. For the Gaussian likelihood, we use an interative coordinate descent approach, where Equation~\ref{eq:latent_update} is used to update the latent transforms, and a quasi-Newton method (L-BFGS) is used to update the model parameters. In the \textit{Laplacian} likelihood,  we solve the linear program in Section~\ref{sec:laplacian} with the dual simplex method \citep{lemke1954dual}.

\subsubsection{Summary}
The presented algorithm goes through a sequential number of steps. First, we build the graph and its associated spanning tree by finding correspondent slices between contrasts (Section~\ref{sec:preliminaries}). Then, we compute the specified registrations above (Section~\ref{sec:graph_structure}) using the learning based approach introduced in Section~\ref{sec:registration_networks}. With this information and the likelihood model of choice in hand, we use Equation~\ref{eq:approximate} to solve the inference problem at each voxel in the space of the low-resolution velocity fields. Finally, we upsample the solution with linear interpolation and integrate it with the scaling and squaring technique \citep{arsigny2006log} to yield the final estimate of the latent deformation fields.

The package is written in Python and made publicly available in \url{https://github.com/acasamitjana/3dhirest}. Registration networks are optimised using PyTorch and the linear program is solved using the implementations found in the Gurobi package \url{https://www.gurobi.com/downloads/}.

\section{Experiments and results}
\label{sec:experiments}

\subsection{Data}
To test and evaluate our framework, we use two datasets: one synthetic, which enables evaluation of registration errors with dense ground truth, and one real,  to test the performance of the model on images with real  distortions due to histological processing.

\subsubsection{Synthetic dataset}

The synthetic dataset consists of T1, T2 and FLAIR MRI scans from the training subset ($N=285$) of the Brain Tumor Segmentation (BraTS) dataset \citep{menze2014multimodal}. The T1 images with contrast (``T1c'') are used as  reference volumes, and the T2 and FLAIR images are used as a proxy for two histological contrasts. The BraTS scans are resampled (by the challenge organisers) to 1mm$^3$ isotropic resolution, and we define the $z$ coordinate along the inferior-superior (I-S) direction. We generate synthetic 2D non-linear deformation fields using a grid of control points and B-spline interpolation \citep{prautzsch2002bezier},  independently for each axial slice of the T2 and FLAIR images, in order to mimic geometric distortion due histological processing (further details in Section \ref{sec:experimental_setup} and Figure ~\ref{fig:images_brats} below).

\subsubsection{Real dataset}
\label{sec:real_dataset}
The real dataset is a set of publicly available images from the left hemisphere from a 34-year-old donor, distributed by the Allen institute \citep{ding2016comprehensive} at \url{http://atlas.brain-map.org/}. The dataset includes a multi-echo flash MRI scan acquired on a 7T scanner at 200 $\mu$m resolution, which we use as reference for the 3D histology reconstruction. Two different histological contrasts are also available at sub-$\mu$m in-plane resolution (which we resampled to 250 $\mu$m for convenience):  641 Nissl stained coronal sections with 200 $\mu$m spacing, and  287 coronal sections with 400 $\mu$m spacing and immunostained using parvalbumin. The correspondence between histological sections is found by matching the closest pair, with errors no superior than $0.1$mm. For each section, we generate an associated tissue mask by thresholding and morphological operations. The final stacks of histological images do not have regular spacing due to missing sections, and there are a number of gaps of several mm without any sections at all, which divide the hemisphere into six different slabs. 
%

In order to estimate the linear registration between the reference volume and histological stacks, we first used our previously presented algorithm \citep{tregidgo20203d} to co-register the Nissl sections and the MRI. Once the MRI was linearly aligned to the stack, we used a block-matching algorithm \citep{ourselin2001reconstructing} (as implemented in NiftyReg, \citealt{modat2010fast}) to independently compute an affine transform between each histological section and its corresponding resampled MRI slice. Moreover, using the left hemisphere from the ICBM nonlinear 2009b symmetric atlas \citep{fonov2009unbiased}, we compute the projection from the subject space to MNI coordinates.

Quantitative evaluation for this dataset is carried out using landmarks generated as follows. First, one salient point for each Nissl section (details below) is automatically sampled. Next, a first observer (JEI) marked the equivalent locations on the parvalbumin sections (where available) and on the resampled MRI. This generates a set of 641 pairs of landmarks for Nissl/MRI, and 287 for Nissl/parvalbumin, which can be used for quantitative evaluation. The same observer (JEI) reannotated the landmarks on a different day, for estimation of intra-observer variability. A second observer (AC) also annotated the same landmarks, for estimation of inter-observer variability. 

The reference landmarks on the Nissl sections were sampled in a manner that ensured both salience and uniform spatial distribution. First, we applied a Harris corner detector \citep{harris1988combined} with a low threshold on the quality of the corner (0.001) to every section. Next, we randomly sampled a location on every section with a uniform distribution, and centred on it a Gaussian kernel with standard deviation $\sigma$=20 pixels (i.e., 5 mm) in both $x$ and $y$. We used this spatial Gaussian distribution to modulate (multiply) the scores from the Harris detector, and picked the landmark with the highest score.

\subsection{Metrics}
The synthetic dataset enables dense quantitative evaluation of the 3D reconstruction methods at the pixel level. Let $\bm{\phi}^c_n(\bm{x})$ represent the ground truth 2D deformation field between the reference slice $n$ and histological contrast $c$, and let $\hat{\bm{\phi}}^c_n(\bm{x})$ be the deformation field estimated by an algorithm. To assess the performance of the presented framework, we define different performance metrics.

We first define the \textit{pixel-wise error} as the bivariate deviation of the estimate from the ground truth deformation field at each pixel. Hence, for slice $n$, contrast $c$, and location $x$, we have:
\begin{equation}
    \bm{e}^c_n(\bm{x}) = \bm{\phi}^c_n(\bm{x}) - \hat{\bm{\phi}}^c_n(\bm{x}).
\end{equation}

Based on this error, we define two performance metrics. First, the \textit{intra-slice error}, defined as a global average over all pixels and slices of the module of the pixel-wise error. The metric is computed only within the tissue mask that results from the intersection of the reference and registered slices, resulting in a valid domain $\Omega_n$ for each slice:
\begin{equation}
    E^c_W = \frac{1}{N}\sum_{n=1}^{N} \frac{1}{\vert \Omega_n \vert}\sum_{x \in \Omega_n} \|\bm{e}^c_n(\bm{x})\|
\end{equation}

We also want to measure the consistency across slices, something that the \textit{intra-slice error} doesn't capture. For this purpose, we use a second metric referred to as \textit{inter-slice error}, which measures the error consistency across the direction of the stack. 
Intuitively, the \textit{inter-slice error} measures the smoothness of the reconstruction, by comparing the consistency of the errors across neighbouring slices -- if errors are consistent, the reconstruction is smooth. Its specific definition is:
\begin{align}
    &E_B^c = \nonumber \\
    & \frac{1}{N-1}\sum_{n=1}^{N-1} \frac{1}{\vert \Omega_n \vert}\sum_{x \in \Omega_n} \| \bm{e}^c_n(\bm{x}) - \bm{e}^c_{n+1}(\bm{x}) \|.
\end{align}
These two metrics complement each other in measuring the trade-of between accuracy and smoothness of the recovered 3D volume. 

We compute them not only between the reference volume and the two (surrogate) histological stains (i.e., T2, FLAIR), but also between these two contrasts - which is easily achieved by defining:
$$
\phi_n^{c,c'} = \left(\phi_n^c\right)^{-1} \circ \phi_n^{c'}
$$
$$
\bm{e}^{c,c'}_{n+1}(\bm{x}) = \phi_n^{c,c'} (\bm{x}) - \hat{\phi}_n^{c,c'}(\bm{x})
$$
and use the estimation error to compute $E_B^{c, c'}$ and $E_W^{c, c'}$

\subsection{Experimental setup}
\label{sec:experimental_setup}
To generate the synthetic deformations in the BraTS dataset, we independently deform each axial slice of the  T2 and FLAIR with 2D deformation fields generated as follows. First, normally distributed,  low-resolution deformation fields of size 9$\times$9$\times$2  are generated independently for each slice, subject and modality (T2, FLAIR). Each element in the low-resolution field is an independent Gaussian variable with zero mean and a standard deviation which is constant for each slice, and which is sampled from a uniform distribution $\mathcal{U}[3,7]$ (in pixels). Each deformation field is then resized to the original image size using B-Spline interpolation. We explicitly avoid using velocity fields to prevent imitating the deformation model used in the algorithm. The final volume is built by applying the deformations and resampling with bilinear interpolation.

To measure the robustness of our framework against large registration errors, we introduce outliers in the synthetic dataset. Specifically, we further distort subsets of the T2 and FLAIR slices (2\%, 5\%, 10\% and 20\%, to test scenarios with increasing number of outliers) by applying large random rotations  ($90^\circ$, $180^\circ$, or $270^\circ$), which lead to large errors in the nonlinear registration algorithm. Each modality is distorted independently. Evaluation is carried out only on the undistorted slices.

In order to analyse the impact of the deep learning registration techniques, we consider two additional diffeomorphic registration algorithms:
\begin{itemize}
    \item \textit{NiftyReg (NR)} (\citealt{modat2012parametric}): we use NiftyReg with SVF parameterisation (``--vel" option), control point spacing of 8 pixels, local normalised cross correlation similarity metric and bending energy penalty term. This setup makes the NR model as similar to our deep learning registration framework as possible. 
    \item \textit{Registration Networks} (RegNet): we use learning-based registration networks, trained as reported in Section.~\ref{sec:registration_networks}. 
\end{itemize}

In our experiments, we compare slice-wise registration using these two algorithms, with different versions of our Spanning Tree framework (henceforth, ST) as well as with a state-of-the-art baseline algorithm: 
\begin{itemize}
    \item ST2-L2: single-contrast framework presented in \cite{iglesias2018model} using masks in the graph and a Gaussian likelihood for the registrations. We run the algorithm independently for each histological contrast.
    \item ST2-L1: same as ST2-L2 but with a Laplacian likelihood.
    \item ST3-L2: multi-contrast framework presented in this work (joint registration of contrasts) using masks and a Gaussian distribution for the registrations.
    \item ST3-L1: same as ST3-L2 but with a Laplacian likelihood.

    \item IR: finally, we implement a state-of-the-art registration refinement approach presented in \cite{adler2014histology}, which we will refer to as ``Iterative refinement'' (IR). This algorithm uses a diffeomorphic registration method (in our case, NiftyReg) to find a deformation that simultaneously aligns each histology section to the correspondent MRI slice and their most immediate neighbours, using a coordinate ascent approach. We run  5 iterations over the stack of histology sections.
\end{itemize}

\subsection{Results on synthetic dataset}
\label{sec:synthetic}

\begin{figure*}[!h]
    \centering
    \includegraphics[width=1\linewidth]{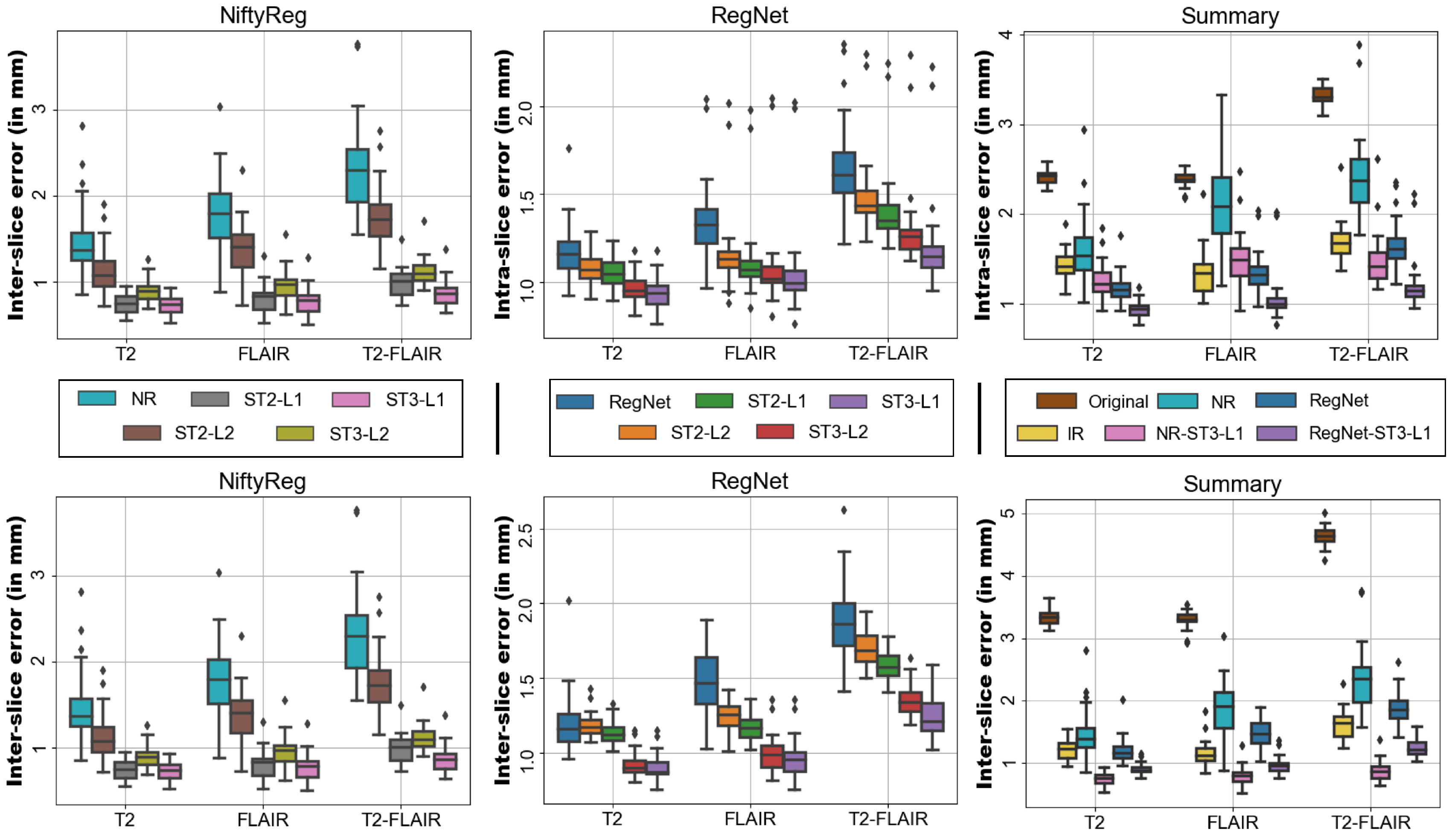}
    \caption{Top row: the first two columns show the \textit{intra-slice} and  \textit{inter-slice} error for NiftyReg and RegNet when used in isolation (i.e., for each slice and contrast independently) and when refined with the different variants of the algorithms proposed in this paper. The last column shows a summary of previous plots comparing the errors for the distorted images, the baseline algorithm (IR), the registration approaches used in isolation, and the proposed approach ST3-L1.
    Each metric is computed for T2 and FLAIR  modalities with respect to the T1, as well as for the consistency error between themselves. }
    \label{fig:methods_brats}
\end{figure*}

In Fig.~\ref{fig:methods_brats} we compare different configurations of the algorithm using NR and RegNet as registration methods. RegNet, which is trained specifically for the problem at hand (as opposed to used a generic optimiser) provides better initial alignment between modalities than NR. Over the initial linear alignment, RegNet improves the \textit{intra-slice} error by 49\% and 42\% for T2 and FLAIR modalities, respectively, compared to the 37\% and 8\% using NR. Moreover, estimation error variability in RegNet is consistently $\sim 5$x times lower than in NR. According to these results, the dependence of standard intermodality registration algorithms (e.g., NiftyReg) on the contrast and image appearance can be partly mitigated using learning-based approaches. Further refinement can be achieved using the presented framework with different modelling options. In our previous work, we used a single-contrast approach with the $\ell_2$-norm as a cost function (ST2-L2), yielding around 10\% extra improvement in each modality. 

The extensions introduced in this work improve upon that result in two ways: \emph{(i)}~the $\ell_1$ norm can correct larger registration errors with respect to direct registrations (no refinement); and \emph{(ii)}~a multi-contrast framework can correct registration errors in one modality using redundant measurements from other modalities. 
We perform statistical significance analysis between all methods compared, with the multi-contrast framework using the $\ell_1$ norm (ST3-L1) significantly outperforming all other algorithms ($p<0.01$), both with NR and RegNet. Over the initial registrations, ST3-L1 yield improvements of 19\% and 22\% (RegNet) and 23\% and 29\% (NR) on the \textit{intra-slice} error, for T2 and FLAIR, respectively. NR yields  smoother reconstructions when combined with our proposed refinement method, while RegNet achieves better inter-modality alignment (Fig.~\ref{fig:methods_brats}, summary). Apart from the quantitative evaluation, the smoothness of the reconstruction is apparent from the visualisation of the error in the orthogonal planes (see Section 2 in the supplementary material). Unless otherwise specified, we use RegNet with ST3-L1 as default configuration for the proposed method throughout the rest of this paper.

\begin{figure*}[!h]
    \centering
    \includegraphics[width=1\linewidth]{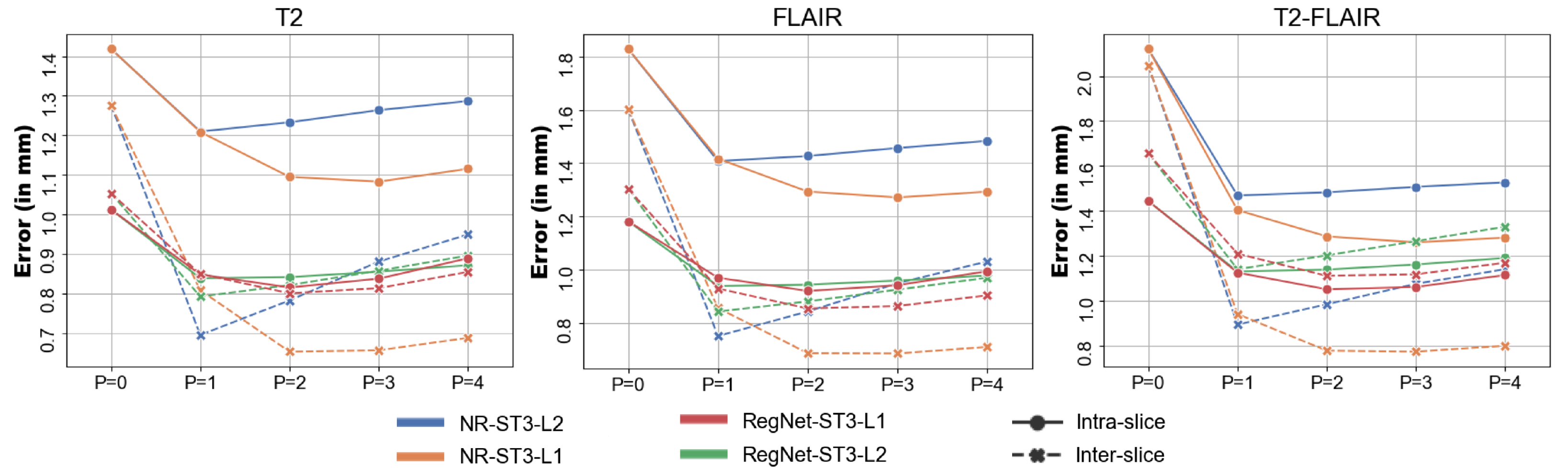}
    \caption{\textit{Inter-} (dashed) and \textit{intra-slice} (solid) errors as a function of neighbours in the observational graph ($P$), for the different versions of our algorithm. Setting $P=0$ is equivalent to running the algorithm independently for each slice.}
    \label{fig:neighbours_brats}
\end{figure*}

Figure \ref{fig:neighbours_brats} shows the results of varying the number of neighbours in the observational graph, $P$. This parameter represent a trade-off between several factors: smoothness, banana effect, z-shift error accumulation, accuracy and computational requirements. For example, the larger the redundancy in the graph (larger $P$) the more robust and smooth the method is at the cost of straighten curved structures. The baseline, $P=0$, builds a disconnected graph using the raw intermodality observations from RegNet directly. The results in this figure show that the optimal number of neighbours depends on the likelihood function. In the $\ell_2$-norm case,  $P=$1 is the optimal number of neighbours for the trade-off between smoothness and banana effect. However, the $\ell_1$-norm is more robust to errors in intramodality registration with increased number of neighbours in the observational graph.  We use a paired t-test to quantify statistically significant mean differences and found that $P=2$ is the optimal number of neighbours for both the \textit{inter-slice} and \textit{intra-slice} error when using the Laplacian likelihood; from $P \geq 3$ the z-shift error start to  accumulate. Therefore, we will use $P=2$ (combined with RegNet and ST3-L1) throughout the rest of this manuscript, unless explicitly specified. 

\begin{figure*}[h]
    \centering
    \includegraphics[width=1\linewidth]{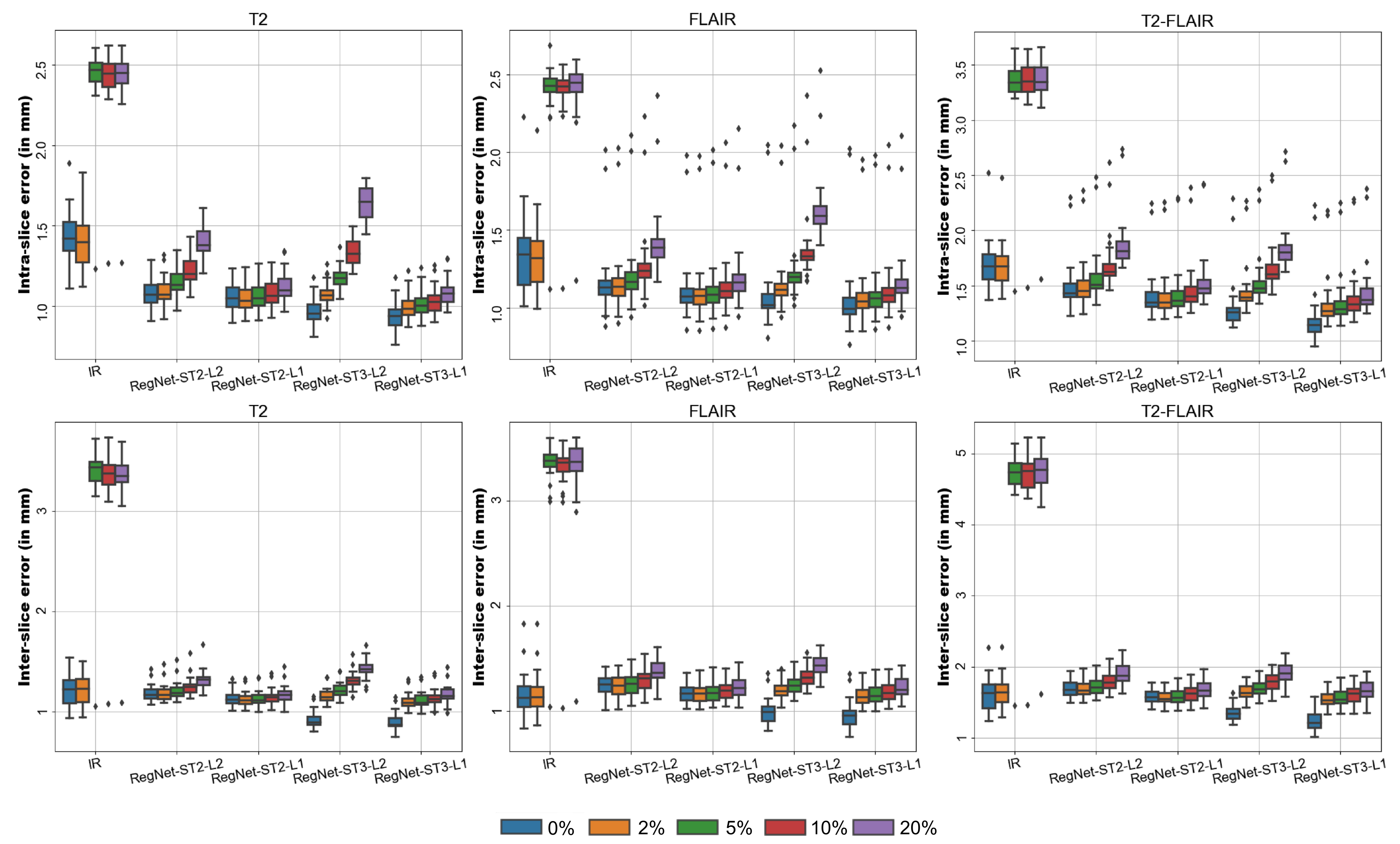}
    \caption{Comparison of the \textit{intra-slice} error (top row) and \textit{inter-slice} error (bottom row) for different level of outliers per modality (0-20\%). The baseline method (IR) and different configurations of the algorithm using RegNet as base algorithm are tested (ST2-L2, ST2-L1, ST3-L2, ST3-L1) and the errors are independently reported for each modality (T2, FLAIR) with respect to the T1, as well as for the consistency error between themselves.}
    \label{fig:outliers_brats}
\end{figure*}

In order to test the robustness of the framework against outliers, we used RegNet for the initial registrations and compared different algorithm configurations against increasing rates of outliers. Fig.~\ref{fig:outliers_brats}  shows the \textit{inter-} and \textit{intra-slice}  error for each contrast as well as the error consistency across contrasts. The presented framework (specially using the $\ell_1$-norm) is more robust than the baseline IR method, which breaks down for more than $5\%$ of outliers. Errors grow with the proportion of outliers, as expected, but the rate at which  performance decreases is different for the different configurations. The $\ell_1$-norm appears to be robust to a considerable number of outliers (up to 10-20\%). The $\ell_2$-norm is much more sensitive to outliers and its performance decreases much faster. We note that the higher error of the ST3 configurations is partly due to the fact that the higher number of registrations $K$ means that more outliers are present in the observational graph. 
Nonetheless, ST3 appears to improve the error consistency and, when combined with the robust $\ell_1$-norm, it boosts the performance of the algorithm.

\begin{figure*}[h]
\centerline{\includegraphics[width=1\linewidth]{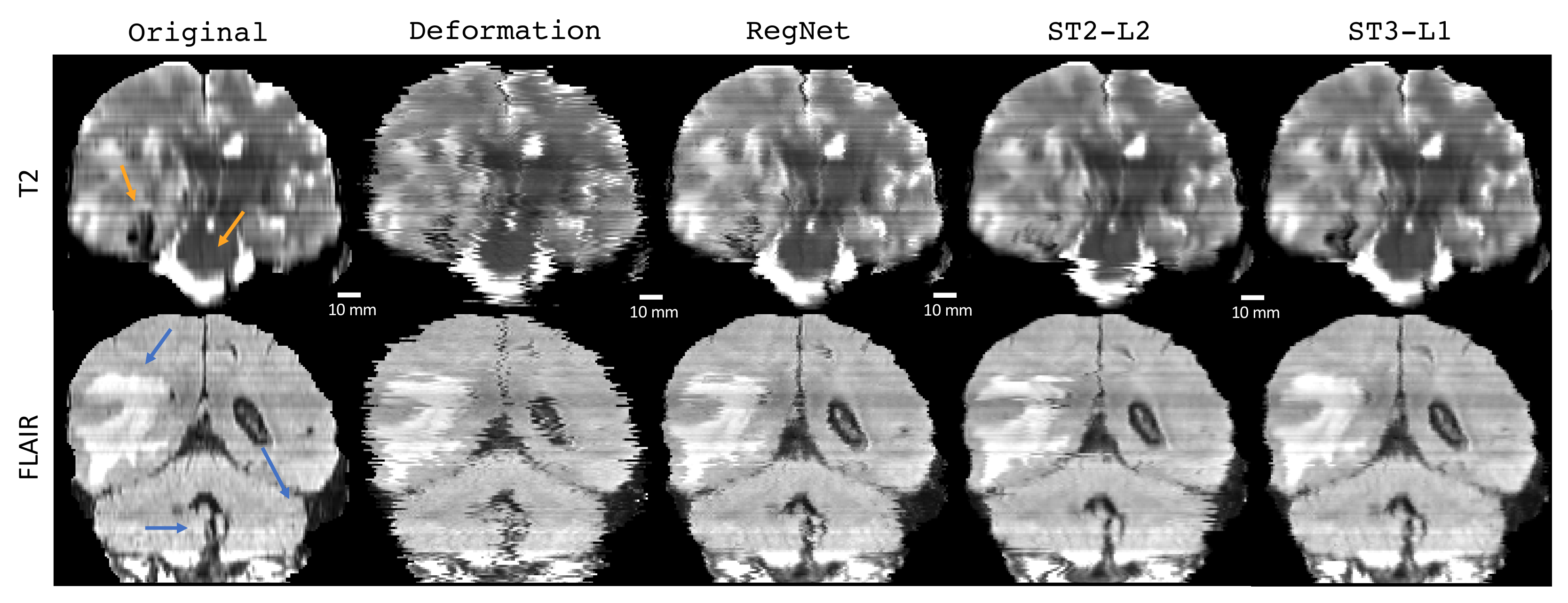}}
    \caption{Reconstructed coronal view for the T2 and FLAIR scans of a sample subject from BraTS \citep{menze2014multimodal}. From left to right we show the original image, the original image with synthetic deformations using B-Splines, the initial registrations using RegNet, the output of the ST2-L2 / NiftyReg algorithm presented in \cite{iglesias2018model},  the output of the ST3-L1 (RegNet)  algorithm presented here, and the ground truth for each contrast. The arrows point at tumorous areas and other regions (e.g., brain stem, cerebellum) which are shown to be hard to register.}
    \label{fig:images_brats}
\end{figure*}

To sum up, the results on the synthetically deformed data show that  ST3-L1 with RegNet  achieves the highest consistency across contrasts and presents the best trade-off between the different factors such as accuracy and smoothness. It outperforms our previous method from \cite{iglesias2018model}, which is based on  NiftyReg and ST2-L2. Qualitative results comparing both algorithms on a case from BraTS are shown in Fig.~\ref{fig:images_brats}. 

\subsection{Results on Allen atlas}
\label{sec:real}

\begin{figure*}[!h]
    \centering
    \includegraphics[width=\linewidth]{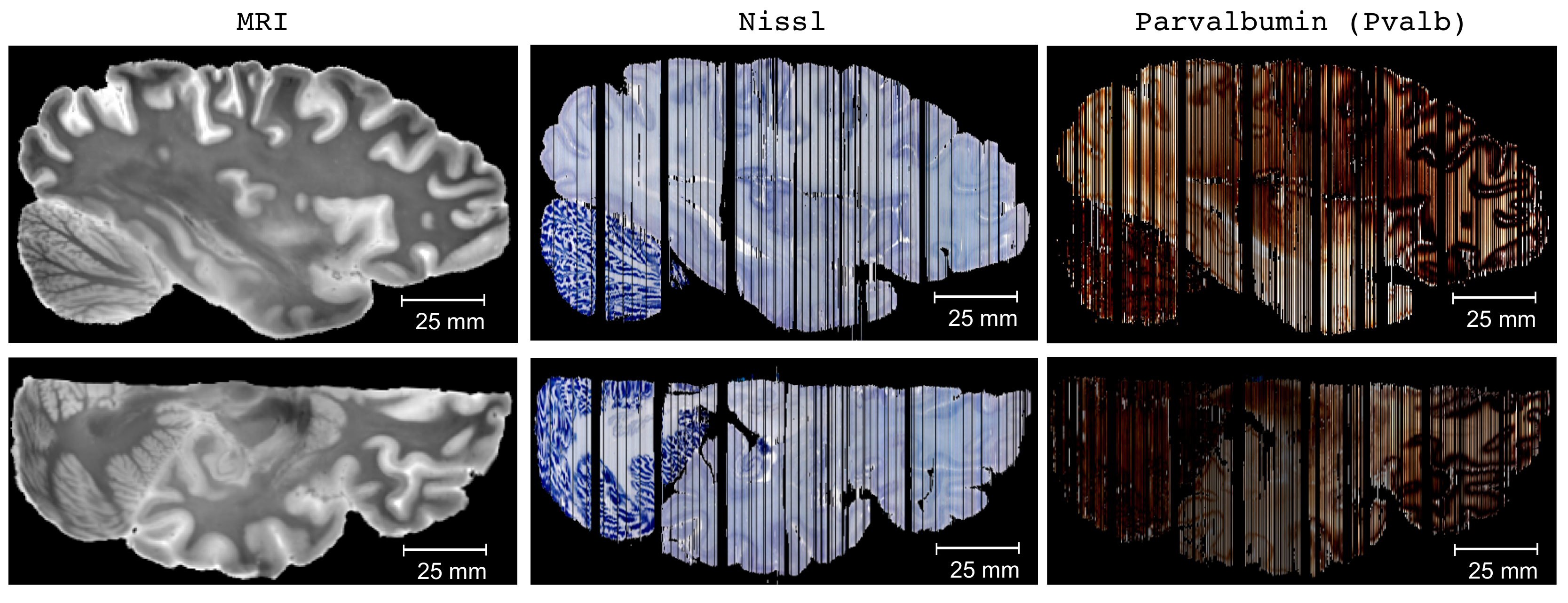}
    \caption{Sagittal and axial views of reference MRI and the reconstruction of the available contrasts (Nissl and parvalbumin), using RegNet combined with ST3-L1.}
    \label{fig:allen_overview}
\end{figure*}

\begin{figure*}[!h]
    \centering
    \includegraphics[width=0.9\linewidth]{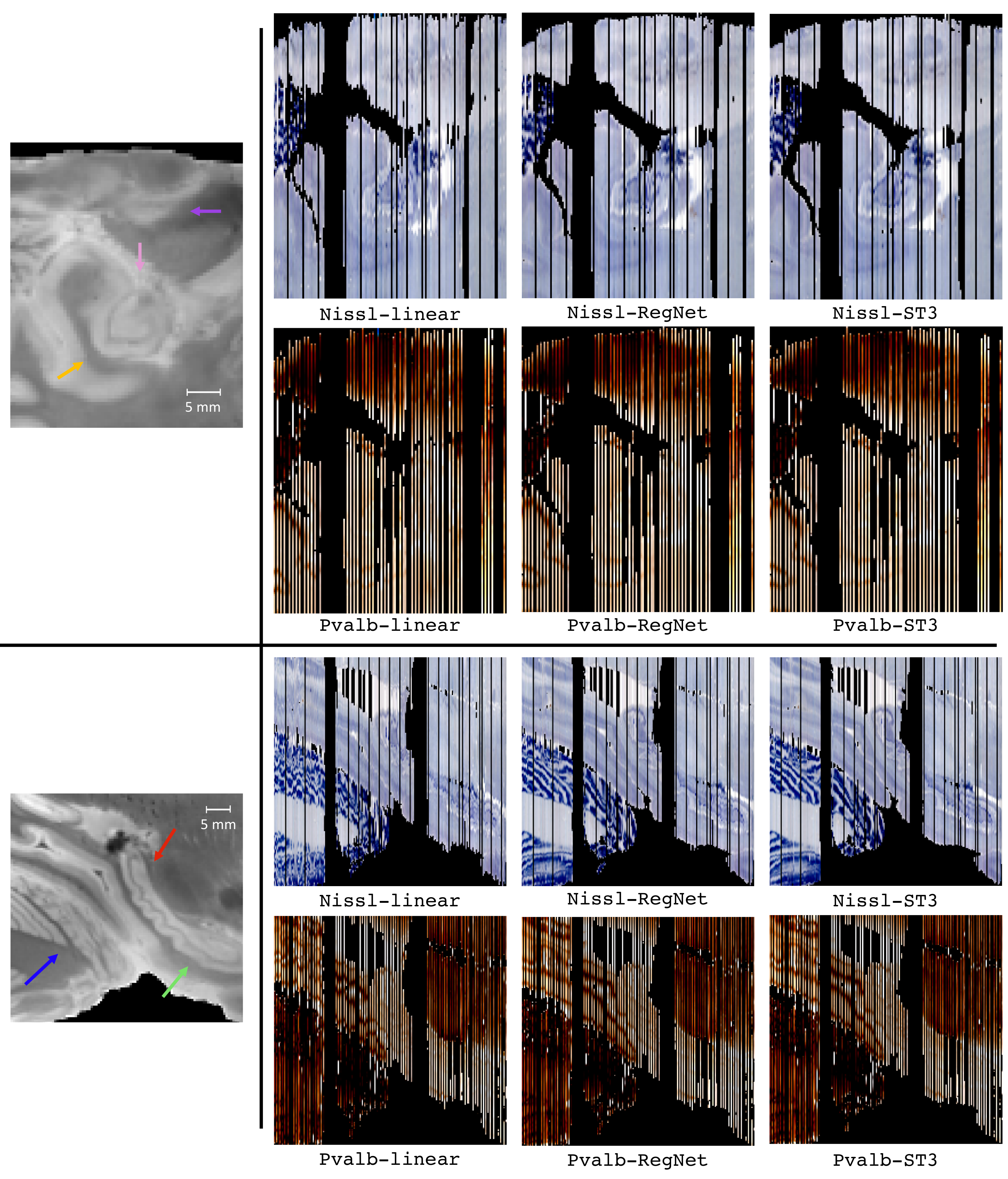}
    \caption{Close-up of 3D reconstructions produced by linear alignment, RegNet alone and its combination with the proposed approach (ST3-L1). Heterogeneous regions such as the cerebellum (red and magenta arrow) or the hippocampus (green and yellow arrows) as well as tissue boundaries (blue and pink arrows) are corrected by the algorithm.}
    \label{fig:allen_closeup}
\end{figure*}

For 3D reconstruction of the  histology of the Allen atlas, we use RegNet as the registration algorithm and ST3-L1 configuration with $P=4$ neighbours in each of the three stacks of images; we increase $P$ with respect to the BraTS dataset to compensate for the lower spacing between sections. Due to a number of larger gaps between sections in this dataset, this process yields a disconnected graph with 5 separate slabs. Therefore, each slab can be  processed independently, which reduces the memory footprint of the algorithm. 

Qualitative results in the sagittal and axial planes are shown in Figures~\ref{fig:allen_overview} and~\ref{fig:allen_closeup}, where the reconstructed volumes  are resampled at 0.25mm isotropic resolution. 3D histology stacks of both Nissl and parvalbumin contrast appear to be aligned with the reference volume at the same time that provide smooth reconstructions. A close-up in the temporal lobe shows that, even though RegNet provides good initial alignment in more homogeneous areas (e.g., cerebral cortex), more jagged reconstructions are recovered in heterogeneous areas such as the hippocampus and the cerebellum. The framework presented here is able to smooth out some of these effects 

\begin{figure*}[h]
    \centering
    \includegraphics[width=\linewidth]{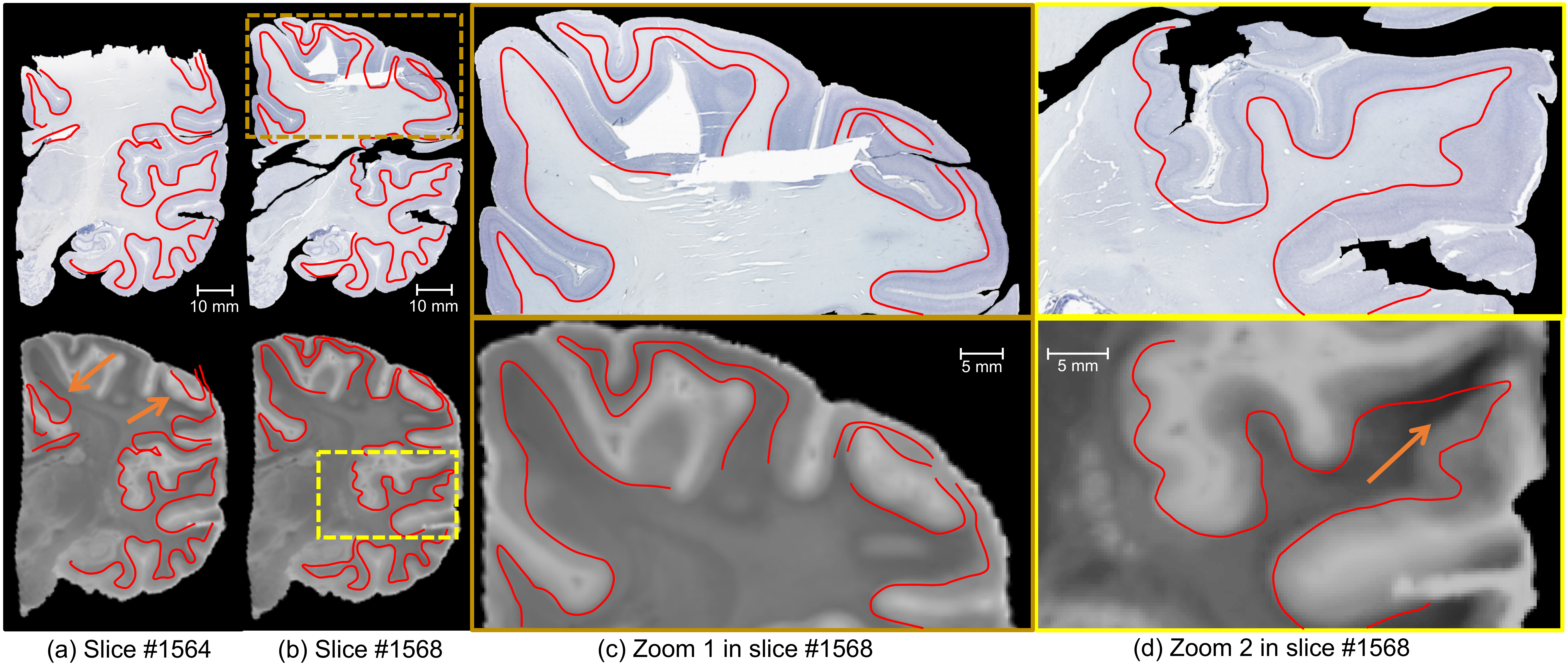}
    \caption{Coronal  view of Nissl staining and MRI for two consecutive sections (number 1564 in (a), number 1568 in (b)), in the presence of artefacts, at high-resolution (8$\mu$m).  The closeup in (c) shows a well registered area from section 1568 with large registration errors in section 1564 due to severe artefacts (missing tissue),  while the closeup in (d) focuses on a region with typical histology artefacts (cracks, holes). We have manually traced the white matter surface  in the histology and displayed it on the registered MRI. The images  show that the proposed method is not only robust but also yields accurate reconstructions.  Arrows indicate large registration errors due to data artefacts, which are neither propagated to the rest of the image nor to the neighbouring sections. }
    \label{fig:allen_coronal}
\end{figure*}

The algorithm  not only recovers smooth reconstructions, but also produces registrations that are accurate and robust. Figure~\ref{fig:allen_coronal} shows the reconstruction of two consecutive heavily damaged coronal sections in plane, i.e., in coronal view. Despite the presence of strong artefacts (e.g., cracks, torn and missing tissue), our method is able to produce smooth reconstructions that are robust against these artefacts and yield accurate 3D reconstructions. Particularly, large registration errors (Figure~\ref{fig:allen_coronal}.a) are not propagated to neighbouring sections (Figure~\ref{fig:allen_coronal}.bc). Moreover, despite the fact that reconstruction is not accurate around histology artefacts (Figure~\ref{fig:allen_coronal}.d), it can be seen that accurate accuracy is preserved  away from the regions with artefacts. The correspondence between the three contrasts on a section with severe artefacts as well as the smoothness of the deformation field in plane is illustrated in Figure 4 in the supplementary material.

Finally, Figure~\ref{fig:allen_boxplot} shows boxplots for the intra-slice errors computed from the manually placed landmarks, and compares them to the inter- and intra-observer variabilities. Even though the median error does not improve with respect to RegNet (as it did on the synthetic dataset, Figure~\ref{fig:images_brats}), there is a small decrease in outliers (specially using NR) after refinement of the registrations with our proposed approach. This is despite the fact that the additional smoothness imposed by our approach (apparent from Figures \ref{fig:allen_overview} and \ref{fig:allen_closeup}) necessarily represents a trade-off with accuracy. Compared with the initial affine alignment, our method achieves reductions of 19\% in Nissl/MR registration, 21\% in parvalbumin/MRI registration, and 30\% in Nissl/parvalbumin in the median error. While these differences may not seem large at first, they have a very noticeable impact on the quality of the output (again, see Figures \ref{fig:allen_overview} and \ref{fig:allen_closeup}).

In absolute terms, our approach achieves median registration errors of approximately 1 mm or below: 0.88 mm for Nissl/MRI, 1.05 mm for parvalbumin/MRI, and 0.83 mm for Nissl/parvalbumin. These are approximately within 0.5 mm of the intra- and inter-observer variability (0.71/0.79 mm for Nissl/MRI, and 0.59/0.73 mm for Nissl/parvalbumin). Using NiftyReg as a base algorithm, landmark misalignment is increased mostly due to errors in the initial observations, with the exception of the Nissl/parvalbumin registration that achieves an inter-observer level of error.

\begin{figure*}[!h]
    \centering
    \includegraphics[width=0.9\linewidth]{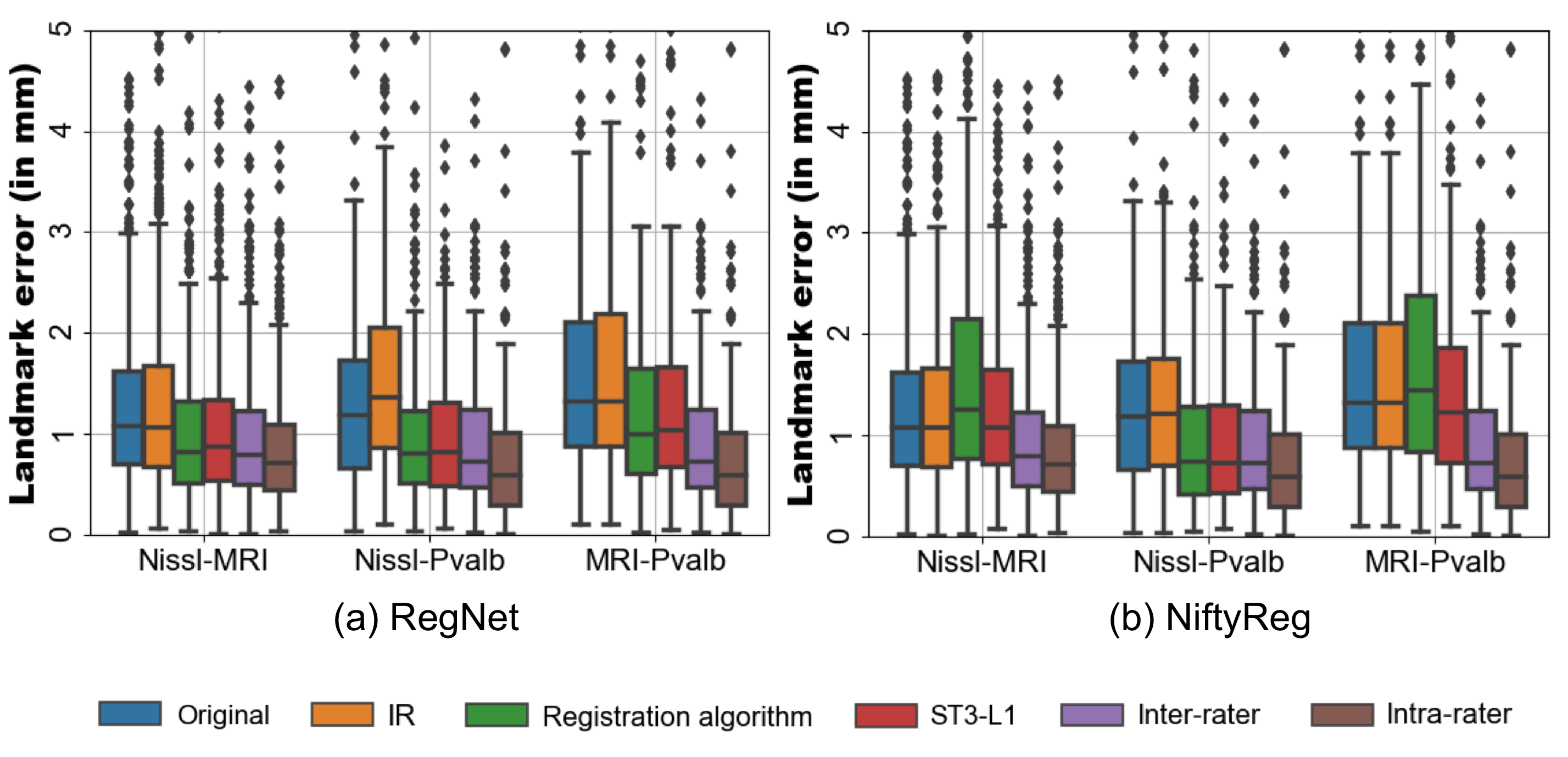}
    \caption{Landmark error using RegNet (left) and NiftyReg (right) as base algorithms for alignment between the Nissl/MRI, Nissl/parvalbumin and MRI/parvalbumin. Inter-slice and intra-slice errors are computed as stated in Section~\ref{sec:real_dataset}.}
    \label{fig:allen_boxplot}
\end{figure*}

\section{Discussion and conclusion}
\label{sec:discussion}

In this manuscript, we have presented a probabilistic model for joint registration of image stacks in histology and MRI. Volumetric histology reconstruction is posed as a Bayesian inference problem, which can be solved analytically ($\ell_2$-norm) or with standard linear programming techniques ($\ell_1$-norm). The algorithm effectively balances banana effect, z-shift effect and registration accuracy to produce 3D reconstruction that are both smooth and precise -- and also robust, if the version with the $\ell_1$-norm is used. Correcting other histology artefacts (e.g., tears or cracks), typically characterised by sharp transitions, is out of the scope of this work, even though our robust framework minimises their impact on the final reconstruction.


A number of assumptions on the data  are made in this work. First, we assume smooth transitions between nodes in the graph. This smoothness depends directly of the regularity of the anatomy and the spacing between sections. We have presented positive results on the Allen datasets, which has spacings (200/400$\mu$m) that are comparable if not larger than those typically used in human brain atlasing (e.g.,~\cite{amunts2013bigbrain,iglesias2015computational,yushkevich2021three}). Therefore, we are convinced that the smoothness assumption will be met but most datasets. We also assume the availability of an external 3D reference like MRI, which is typically acquired in most 3D histology works to avoid z-shift and banana effect~\cite{malandain2004fusion}. This reference volume is linearly aligned to the stack of histology sections, a problem that is much simpler than estimating the in-plane non-linear deformations; iterating between a 3D rigid body transform and the 2D non linear refinement yields very fast convergence of the 3D transform. Finally, we also assume the availability of tissue masks, that are used to reduce the impact of remaining histology artefacts and to reduce the complexity of the  problem in inference. In practice, these masks can be easily computed using simple thresholding and morphological operations thanks to the strong contrast between tissue and the (glass) background.

Our algorithm builds on standard registration techniques, with the only requirement that they parameterise the deformations with stationary velocity fields. This parameterisation and the likelihood models presented here enable us to define a convex problem with a unique global optimum. Nonetheless, the truncation of the BCH formula, despite not being uncommon in the literature \cite{lorenzi2014efficient,rohe2016barycentric,sivera2019model}, introduces some error in the calculations. Yet, this error is generally quite small (see results in Section 1 of the supplementary material) and is outweighed by the benefits in terms of optimisation, i.e., the guarantee that the global optimum will be found.

In our experiments, we have compared a classical algorithm building on explicit optimisation (NiftyReg) and an unsupervised machine learning approach building on modern neural networks (RegNet). The latter learns a global, data-dependent deformation model that, in practice, is shown to be more accurate and superior to generic optimisation algorithms such as the conjugate gradient strategy used by NiftyReg. While the neural network requires (unsupervised) training, it provides quick predictions for the measurements $\{\mathcal{R}_k(\bm{x})\}$ (i.e., the pairwise registrations), which can be computed in the order of seconds to a few minutes, depending on the size of the problem (number of images, number of contrasts, size of the images, etc.). NiftyReg is between one and two orders of magnitude slower, as it iteratively optimises the deformation for every $\mathcal{R}_k$, such that computing the whole set of registrations take between minutes and hours. Such differences are reduced when accounting for network training times, being more efficient for larger datasets e.g., several minutes per subjects in the BraTS dataset. These computed registrations are used to build a connected graph (i.e. all nodes are connected by at least one path) defined by the underlying spanning tree;  otherwise, if some nodes remain isolated, the solution would be ambiguous and extra regularisation would be required for inference. Spatially varying subgraphs of the initial graph enables faster (avoid regions far from tissue) and more robust inference (remove observations with artefacts, such as missing tissue or holes). Once the registrations have been computed, solving the linear program at the control points takes approximately 20 seconds for a BraTS case on an Intel Core i7-9800X processor with 8 cores (150 minutes if the problem is solved at every pixel instead). 


The extension to multiple contrasts has two main implications. First, there is an increasing demand of computation and memory, as the number of observations increase quadratically with the number of contrasts. And second, there is an increase of the number of cycles in the observation graph, making it more redundant and robust. In other words: since the number of latent variables increase linearly (rather than quadratically), the inference model becomes progressively more overdetermined. 
Moreover, a multi-contrast framework is shown to improve the consistency between reconstructed volumes for different histological contrasts as seen in Fig.~\ref{fig:methods_brats}. The use of an $\ell_1$-norm further increases the robustness of the framework in the presence of low to moderate levels of outliers. No running time difference has been found in practice between using an ST3 approach or solving the inference problem independently for each contrast using ST2. Overall, we showed that our framework provides a more robust and accurate alternative to the baseline iterative method IR.

The usefulness of the framework we have presented has been shown with a publicly available real test case from the Allen human brain atlas, which has two available contrasts (stains). Despite the fact that the two stains are sampled at different frequencies (yielding an irregular graph structure) and that typical histology artefacts are found in both contrasts (e.g., tears, folding, cracks, inhomogeneous staining), our method is able to recover the original 3D shape  with smooth transitions between slices.  We have shared our results through the OpenNeuro repository (https://openneuro.org/datasets/ds003590) along with the mapping to MNI space, providing the neuroimaging community with a cross-scale link between the two atlases.

In the modelling step, different assumptions are made about the observed deformation fields, such as the independence between registration noise and spatial location, or the conditional independence of the observations given the latent variables. While these assumptions are violated to different extents in different scenarios (application, base registration algorithm), the results on the two datasets have shown that our proposed method works well in practice.  We speculate that the different degrees of departure from the assumptions may be behind some of our empirical results. 

Extensions of this work can follow several directions, being the first one evaluating learning-based registration-by-synthesis approaches to improve initial intermodality alignment \citep{qin2019unsupervised,xu2020adversarial}. A second direction is to consider a more realistic approach to model the registration errors that accounts for spatial correlations. A third step would be to integrate intensity homogenisation techniques in the framework to jointly correct for uneven staining. Moreover, imputation methods to fill the gaps and improve continuity between sections may be very well explored. 

We plan to use this framework on the 3D histology reconstruction pipeline introduced by \cite{mancini2020multimodal} to build a $\mu$m-resolution probabilistic atlas of the human brain. The accurate 3D reconstruction of histological atlases with the proposed method will enable volumetric studies of the whole human brain at the subregion level, with much higher specificity than current approaches.

\section*{Acknowledgement}

This work was primarily funded by the European Research Council (Starting Grant 677697, project ``BUNGEE-TOOLS''). %
SF is supported by the EPSRC-funded UCL Centre for Doctoral Training in Medical Imaging (EP/L016478/1) and Doctoral Training Grant (EP/M506448/1). %
Support for this research was provided in part by the BRAIN Initiative (1RF1MH123195-01, U01MH117023), the National Institute for Biomedical Imaging and Bioengineering (P41EB015896, 1R01EB023281, R01EB006758, R21EB018907, R01EB019956, P41EB030006), the National Institute on Aging (1R01AG070988-01, 1R56AG064027, 1R01AG064027, 5R01AG008122, R01AG016495), the National Institute of Mental Health (R01 MH123195, R01 MH121885, 1RF1MH123195), the National Institute for Neurological Disorders and Stroke (R01NS0525851, R21NS072652, R01NS070963, R01NS083534, 5U01NS086625,5U24NS10059103, R01NS105820), and was made possible by the resources provided by Shared Instrumentation Grants 1S10RR023401, 1S10RR019307, and 1S10RR023043. Additional support was provided by the NIH Blueprint for Neuroscience Research (5U01-MH093765), part of the multi-institutional Human Connectome Project. In addition, BF has a financial interest in CorticoMetrics, a company whose medical pursuits focus on brain imaging and measurement  technologies. BF's interests were reviewed and are managed by Massachusetts General Hospital and Partners HealthCare in accordance with their conflict of interest policies. %
This work was also partly supported by core funding from the Wellcome/EPSRC Centre for Medical Engineering [WT203148/Z/16/Z; NS/A000049/1]. %
Additional support was provided by Alzheimer's Research UK (ARUK-IRG2019A-003) and the NIH (1RF1MH123195-01, 1R01AG070988-01).

\bibliography{mybibfile}

\end{document}